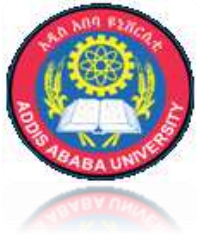

ADDIS ABABA UNIVERSITY

SCHOOL OF GRADUATE STUDIES

COLLEGE OF NATURAL SCIENCES

DEPARTMENT OF COMPUTER SCIENCE

A Framework for Multi-source Prefetching Through Adaptive Weight

By

Yoseph Berhanu Alebachew

A thesis submitted to the school of Graduate studies of Addis Ababa University in partial fulfilment of the requirements for the Degree of Master of Science in Computer Science

May 26, 2014

ADDIS ABABA UNIVERSITY

SCHOOL OF GRADUATE STUDIES

COLLEGE OF NATURAL SCIENCES

DEPARTMENT OF COMPUTER SCIENCE

A Framework for Multi-source Prefetching Through Adaptive Weight

Yoseph Berhanu Alebachew

Advisor: Dr. Mulugeta Libsie

Name and Signature of members of the Examining Board:

| Name | Signature |
|---|---|
| 1. Mulugeta Libsie (PhD) | \_\_\_\_\_\_\_\_\_\_\_\_\_\_\_\_ |
| 2. Solomon Atnafu (PhD) | \_\_\_\_\_\_\_\_\_\_\_\_\_\_\_\_ |

## Table of Contents

# List of Figures

# Acknowledgment

First and for most I thank God for the life and achievements I had so far and for what is to come, it has been nothing but a bliss. His love for me was expressed to me when he gave me my parents without whom this all would have been just a dream. Thank you my parents for your unparalleled support. I would also like to express my deepest gratitude to my advisor Dr. Mulugeta Libsie for his supreme encouragement, guidance, and support.

My colleagues, especially, Dessalegn Mequanint, Fasika Minda, Seble Nigusse, Yonathan Getachew, Mohamood Abdela and others share a great deal of my gratitude for their continuous provision and baking. Finally I would like to thank the legion of teachers and mentors I was fortunate enough to come across, who never failed set the canon of success a bit higher for me which eventually lead to this thesis. Thank you for everything.

# Abstract

The World Wide Web has come to be a great part of our daily life, yet user observed latency is still a problem that needs a proper means of handling. Even though earlier attempts focused on caching as the chief solution to tackling this issue, its success was extremely limited. Prefetching has come to be the primary technique in supplementing caching towards soothing the latency problem associated with the contemporary Internet.

However, existing approaches in prefetching are extremely limited in their ability to employ application level web document relationship which is often visible only to the content developer. This is because most approaches are access history based schemes that make future users' access prediction only based on past user access. Attempts to incorporate prefetching schemes that utilize semantic information with those that use users past access history are extremely limited in their extensibility.

In this work we present a novel framework that enables integration of schemes from both worlds of prefetching (i.e., history based and semantic schemes) without the need for a major modification to the algorithms. When there is a need/possibility to capture new application level context, a new algorithm could be developed to do so and then it can be integrated into the framework.

Since each participating scheme is merely viewed as an algorithm that produces a list of candidate objects that are likely to be accessed in the near future, the framework can entertain any one of the existing prefetching schemes. With its adaptive weight management technique the framework adjusts the effect of each algorithm in the overall prediction to parallel with its observed performance so far.

We have found this formwork to be less aggressive than its contemporary counterparts which is extremely important for resource constrained mobile devices that have come to be the major means of access by users of the current web.

# Chapter 1 - Introduction

## 1.1 Background

As technological advancements have been made, the speed of network connections have escalated in an encouraging manner. This fact, together with reduced cost of communication, opened the door for more interesting and complicated applications to be developed [1], which in turn led to the rise of data items with huge size and variation like never seen before. Hence, the latency experienced by applications (subsequently by users) in accessing these data is still a problem that should be dealt with.

In parallel, from their advent, computer networks have transformed our lives in such a way that it is now almost impossible to think of our daily ritual without them. Nowadays everything we do involves these networks [2, 3, 4, 5, 6, 7, 8]. Starting from how we socialize to how we do business and even to the point of getting health assistance and medical treatments relies on computer networks.

This growth in our dependency towards computer networks calls, at the very least, for its reliability and efficiency, if not more. The most important direction in terms of efficiency for a computer network is its speed and ability to hide latency from users. This further underlines the need for techniques that deal with latency.

Fang et al. [9] outlines five potential sources of latency, namely, web servers' heavy load, network congestion, low bandwidth, propagation delay, and bandwidth underutilization. That said, one obvious solution for reducing latency is increasing bandwidth. However, this would only be the naïve approach towards solving the problem at hand due to two facts, especially for the case of Wide Area Networks (WANs). These are cost associated with improving bandwidth capacity and the very basic characteristic of Mother Nature.

As one could expect, upgrading network bandwidth is costly. Furthermore, it is impossible to rely on these improvements as a definitive answer for such impediments as physical distances will always exhibit its influence on data transmission speed [10]. Even if we are tempted to ignore this fact and aim to consider such improvements as core solution towards reducing latency, it can only be projected that higher bandwidth would ease users' endeavour to create more sophisticated applications that produce "heavy" data.

Putting these facts in mind, techniques such as caching and prefetching have been developed, each in plethora of variations. Caching refers to the storage of recently retrieved objects in a spatially closer location to the user in an attempt to enable fast access in the near future [1, 11]. Caching by itself, however, is extremely limited in the level of improvements it provides in terms of latency hiding. A typical caching policy with an infinite storage size is limited to a maximum hit ratio of 40%-50% [12, 13]. The main reason for this is the tendency of users often to request a new object as the time goes by rather than asking objects they have viewed earlier.

To improve the performance of caching, they are accompanied with prefetching techniques which improves the performance by up to 60% compared to the 26% improvement obtained by a caching technique implemented solely [14, 15]. Prefetching is an attempt to pre-retrieve an object to cache before it is requested by the user based on a systematic guess of which objects will be requested in the near future. This is feasible only if there is an intelligent component that could come up with an accurate decision on what the user will request. The systematic guess is also known as user future access prediction and the component within the prefetching scheme responsible for this task is called the predictor. User future access prediction is not an exclusive challenge in the arena of prefetching. It is also applicable and plays an important role in domains such as user profiling, recommender systems, adoptive websites and so many others [16].

Walled et al. [11] classifies prefetching approaches into two, based on the data used for prediction: history based (i.e., those that take access history for prediction) and content based prefetching (i.e., those that use the content of the current request to find a semantic relationship that could serve as a hint in anticipating future requests). However, a major share of the work done is focused towards user access history based approaches. This is mainly because of the scarcity of applications that present semantic information in a standard manner which leads to a relatively inferior performance of content based approaches in hiding latency.

However, through time, both the availability of semantic data and our ability to efficiently and effectively process it are making encouraging progress [17]. Commercial companies like Google, Yahoo and Bing along with standardization organizations such as W3C are standardizing the way semantic information is represented on the Web [18, 19]. These same companies and others such as WolframAlpha, Amazon, Swoogle, Yummly, Thinkglue, Hakia, and many governmental organizations are making use of and/or presenting data with semantic relationship depicted in one of the industry standards like RDF and OWL [20, 21, 22].

Even though history based approaches currently seem to outperform their content based counterparts, both approaches possess their own drawbacks. Some of these drawbacks are common to both techniques. For example, the performance of both of these approaches is highly dependent on the location of their implementation in the network architecture (i.e., client, proxy or server). What is more, both of this schemes fail to obtain the help of the other in a transparent manner (i.e., history based approaches supplemented by semantic relation based approaches and vice versa). As a result, history based approaches fail to utilize semantic relationship even when it is available with a high level of accuracy.

This inability to share information together with the high level of dependence of the location of implementation leads to algorithms that work on a partial view of the entire system. On the Internet architecture stated above different information is available at different locations. Servers oftentimes know more about the content while clients know more about the user. Proxy servers are in a position to view a user as part of a group, often composed of users with similar or related interest, concurrently they (i.e., proxies) are able to gauge servers' popularity specifically in a group of users, not just any random set of users on the Internet [23, 24, 25].

Though limited, some attempts were made to combine content based approaches with semantic based ones, and two or more history based approaches together. This class of approaches is referred to as multi-source prefetching. Formally defined, multi-source prefetching is a form of prefetching that combines prediction from multiple sources [26]. A source in this case is any one of the plethora of contemporary algorithms described in the literature review and related work Chapters. In this regards, Prediction-by-Partial Matching (PPM) or Multi-order Markov (discussed latter) model could be perceived as a multi-source scheme since it combines predictions from multiple models.

Most of the works on multi-source prefetching that combined content and history based algorithm constructed the scheme by prioritizing towards history based algorithms. These works created a fixed size bucket of possible objects to prefetch and tried to fill this bucket via one or more history based scheme(s). When these history based schemes fail to come up with enough number of objects to suffuse the bucket, it is only then that the content based approaches are consulted to address this shortage.

However, Davison [26], in particular, attempted to create a scheme that combines multiple algorithms each with a different view point of the overall web data to create a collaborative algorithm. His work did not favour any algorithm while making a predication and subsequently filling the bucket, rather it considers all of them to be of equal precedence. This work assumes that all algorithms are equally capable of predicting the next user access at **all time** without an exception. This is reflected by setting the weight of each algorithm to be of equal value all the time. But, with this assumption the obvious fact that some algorithms are in a better position to know a lot about a certain context than the others is completely ignored.

In this work we propose adaptive weight to be one solution in this direction. We present a framework for multi-source prefetching with adaptive weight that enables existing algorithms to collaborate with each other with the aim of achieving an improved overall performance. With this framework, the overall scheme can make use of information relevant to prediction of next user access regardless of where it is located or which algorithm generated it. In this framework any of the multitude of algorithms available could be implemented where it performs best.

Via a dynamic management of algorithms' weight and the ability to virtually turn on and off a scheme as the user enters a specific domain, the framework can incorporate semantic relationship based prefetching algorithms without the need for manual user intervention. This way when a more domain specific algorithm is available, it can be included into the scheme without the user even knowing.

## 1.2 Problem Statement

Existing prefetching schemes fail to combine predications by multiple algorithms, which have different view point, in a dynamic and an algorithm independent manner.

**Research Hypothesis**

The use of "Multi-Source Prefetching with adaptive weights" is possible and it can enhance the overall performance of existing prefetching schemes in terms of effectiveness and efficiency.

**Research Questions**

The following research questions are going to be investigated:

- How best will the use of multi-source prefetching with adaptive weights improve the predictive performance of prefetching schemes?
- How best will the use of multi-source prefetching with adaptive weights address the issues that arise from changes in the cyber world trends, in particular changes in user end device types?

## 1.3 Motivation

Both the mechanism and motivation of accessing the Internet has dramatically changed in the past decade. The era of personal computers has more or less given a way to the period of mobile devices which in turn seem to be taken over by the age of wearable computers. The latter two present resource constraints which highly affect the user experience. These constraints range from limitation in memory to bandwidth size [27, 28].

What is more, these devices are expected to integrate to the daily life activities of their users seamlessly (i.e., with minimal to no distraction) for them to be acceptable. In contrast to personal computers where users intentionally sit next to a computer often times with the sole purpose of using a computer, mobile and wearable devices are meant to integrate into one's life and are expected to be functioning even when the user is not interacting with them deliberately per se.

With regards to the motivation for accessing the Internet, users now perform tasks that lay at the very fabric of their personal, professional and social lives using the Internet. In its infancy the Internet was a means to communicate simple messages via chat rooms, emails and newsgroups. However, this simplistic view is a past history as users utilize the Internet to undertake a sizeable share of their daily task. We now use the Internet to watch movies [29, 30], buy goods and services [31], access our bank and financial information [32, 33, 34], form social and business gathering in virtual meeting rooms, attend classrooms conducted on the other side of the globe [35, 36, 37] and so much more.

In parallel, following the rise of cloud computing, we have now reached to a point where client end devices need to have no more than a browser to serve as a fully-fledged personal computer without the need to have even a single application software installed on them. Hence, the Internet is no longer the 'plus to have' tool it is a must to have one. Companies like Google and Microsoft are pushing for thin client applications that heavily rely on the Internet to even perform jobs such as text and spreadsheet processing. Popular desktop application developers like Adobe are now delivering their products via Software As a Service (SAS) technology again via the Internet [38, 39, 40].

The above two facts introduce even more stress on the network we have today. Applications are data hungry like never before. On top of that, even if hardware technology has grown drastically in terms of improving network speed (especially following the introduction of fiber optic cables), bandwidth is still an issue due to the move towards wireless communication medium following the rise of mobile and wearable devices.

Thus, we need to devise new mechanisms to address this issue and we should re-evaluate our approach thus far. In this work we propose collaboration and combination of existing algorithms to be one possible solution. This move towards collaboration has proven to be successful in other problems of computer science such as high performance computing and machine learning in classification problems [41, 42, 43, 44].

Moreover, having more and more semantic information being present on the web we should expect to have new algorithms that exploit this to enrich the user experience through prefetching. As a result we should enable the integration of these new algorithms into the well-studied and stable legion of algorithms we have been making use of for a while.

Furthermore, it is not uncommon for content developers to take actions into their own hands to handle latency hiding. Web developers in the past have incorporated prefetching into their design by pre- retrieving, for example, images in the background in presentation like directed navigations [45, 46, 47]. Such a framework will allow the content developers focus on what they do best - content development. What is required from the developers is to provide a procedure that predicts next access using the semantic relationship they know when developing the content. They do not need to worry about prefetching specific issue like cache integration and cache management.

### 1.4 Objectives

#### 1.4.1 General Objective

The general objective of the research is to design a framework that enables transparent multi-source prefetching to improve the predictive performance prefetching schemes.

#### 1.4.2 Specific Objectives

More specifically, the research aims to achieve the following.

- To design a framework that enables prefetching with multi-source collaboration without a major change to contemporary algorithms or schemes which are actually tributaries to the major framework.
- To develop metrics to gauge the efficiency and effectiveness of the proposed solution or scheme.
- To develop a prototype system in order to validate the proposed prefetching scheme.

## 1.5  Scope and Limitations

The research work only targets the items listed in the objective section. It only addresses issues pertinent to prediction of future user access as it applies to prefetching in client/server or related model where there is a need for remote access to digital data placed in a central repository. As a result issues related to prefetching other than the main component such as user session determination technique will not be explored in detail.

## 1.6  Methods

With the aim of achieving the research objectives outlined above the following methodologies will be applied.
- Conducting deeper analyses of literature or state-of-the-art with respect to prefetching techniques.
- Developing input and output metrics.
- Exploring techniques appropriate for predicting data objects that need to be prefetched.
- Developing a syntactic data generator.
- Implementing a prototype based on the proposed framework for evaluation purpose.

## 1.7  Application of Results

The output or proposed solutions of the research can be implemented or used towards enhancing the response time of various client devices that try to access data from a remote repository. Potential application domains include systems in healthcare, news service, Enterprise Resource Planning (ERP), Customer Relationship Management (CRM), Human Resource Management (HRM), Content Management (CM), Service Desk Management and similar sectors characterized by high level of inclination to SAS paradigm.

## 1.8  Thesis Organization

The rest of this thesis is organised as follows. State of the art and related work in the domain of prefetching are presented in Chapters 2 and 3. The newly proposed framework is presented in Chapter 4 followed by evaluation and findings in Chapter 5. Chapter 6 presents discussion of the results. In Chapter 7 conclusions and future works are presented.

# Chapter 2 -   Literature Review

The success of the World Wide Web (WWW) had and will continue to rely on its speed and responsiveness [48]. Unlike the earlier days whereby the computation world tolerated tasks to take hours if not days, these days, everyone is getting used to a faster pace, be it in terms of processing (i.e., CPU throughput) or data communication. Users no longer have to endure data communication that takes up minutes. This accompanied with the fact that mankind is more or less basing its core functions on the cyber world pushes for an improved Internet, among other things. In this regard, cloud computing is one evidence of the need for upgrading, as users expect an equivalent, if not superior, experience from cloud services compared to contemporary desktop environment.

In the past caching was considered to be the frontrunner solution for soothing problems of delay and poor responsiveness in the networking domain [11]. Caching refers to storing of objects that are likely to be requested in the near future in a location closer to the user. Accordingly, a caching hierarchy is created [49, 50] with, ideally, the most probable item to be accessed soon placed closer to the user. Caching focuses on storing objects that have been accessed already by the users. Hence, objects that were never accessed are virtually invisible to the caching system.

Primarily web caching provides three attractive advantages to the stakeholders in a web transaction, which include users/clients, network managers, and content creators [11, 1]. Caching reduces load on content origin server because a portion of the request is satisfied from the cache (i.e., cache hit), hence, not even reaching the server. Furthermore, caching lessens the user perceived latency by exploiting the spatial proximity of the cache storage to the requesting user compared to that of the origin server. What is more, cache hit prevents the need for requesting traffic to traverse beyond the caching point, hence, decreasing network bandwidth consumption.

With all these advantages in mind, though, caching by itself is extremely limited in the degree of improvement it can provide. The principal reason for such a limitation is the tendency of users to request for new objects which were never accessed in the past in hopes of embarking on fresh information. Thence, even with unlimited storage (i.e., with the ability to store everything that has been accessed in the past), performance of caching is limited to about 40% hit ratio regardless of the caching algorithm used [51].

Furthermore, in earlier days simply increasing the cache size could have showed a significant improvement in performance of caching systems due to the existence of only a limited set of applications and server types. However, a rise in popularity of the web led to an escalation not only in the number of clients and their means of access, but also in the number and variation of servers and the applications they provide, thus leading to the need for a smarter mechanism for improving its performance aside from a naïve attempt to increase the cache storage size [52] (as referenced in [53]).

This calls for an alternative mechanism that could supplement the encouraging enhancement observed via caching; prefetching came into light with this in mind. Prefetching refers to the process of systematically anticipating a user's future access and retrieving objects to cache, hopefully even before the user requests them. Prefetching can improve performance in terms of latency reduction to up to twice of implementing caching alone [15].

The primary task in prefetching is anticipating a user's future access. This task (i.e., future user access prediction), however, is not an exclusive challenge in the arena of prefetching. It is also applicable and plays an important role in domains such as user profiling, recommender systems, adaptive websites and so many others [16]. Hence, it has been centre of attention in those domains as well [54, 55, 56, 57]. Exploitation of the user's access history has been the relatively successful approach toward achieving this task so far [16].

In addition to user access prediction, web prefetching involves the determination of user transaction from a set of requests and cache integration, among other things. Even though, a number of mechanisms have been proposed for session determination [58, 59], since the most important component in prefetching is the prediction algorithm, most of the prefetching schemes introduced so far focus on the prediction algorithm subsequently utilizing more or less similar approaches for session determination and other components [48, 60].

On the other hand, multi-source prefetching refers to a form of prefetching that combines predictions from multiple algorithms [26]. In multi-source prefetching, these multiple algorithms that contribute predictions are what we call sources, hence the name multi-source. This need for combining multiple sources arises from the fact that each of these sources/algorithms has a partial view of what is happening. To state it in other words, each algorithm makes use of a limited portion of the information available to assist in the user future access prediction task. For example, content based schemes make use only semantic information while history based schemes use an algorithm that searches for a pattern in a single access log.

For this work, we refer to any information available for an algorithm to utilize in the prediction process to be a context. Here we adopt a definition of context from [61], which is:-

> *Context is any information that can be used to characterise the situation of an entity. An entity is a person, place, or object that is considered relevant to the interaction between a user and an application, including the user and applications themselves*

In the scope of prefetching entity, from the above definition, could be the prediction decision, hence, context is any information that can be used to conclude what to prefetch. This context information is available in different levels and at different locations. In terms of level we mean the degree and type of context information available may vary from application to application.

Accordingly, multi-source prefetching could also be referred to as multi-context prefetching. As it will be shown in later sections, current prefetching approaches are extremely limited in the amount of context information they put into consideration while coming with a prediction of future access and retrieving objects. Contemporary works only consider either the user access history from a single access log or very basic semantic relationship between web object.

## 2.1 Prefetching Classification

Prefetching approaches can be classified into different categories based on a number of criteria. Different authors use different measures for the classification process as presented below.

a. Based on the information used as an input for the prediction process, Nanopoulos et al. [16] divided prefetching approaches into two broad categories; informed and predictive prefetching. For the case of informed prefetching, clients give some sort of hint or at times indicate the exact object to prefetch to the system. Predictive approaches, on the other hand, analyse the access history to come up with an effective prediction of future access.

   Predictive method of prefetching is also known as history based prefetching as there exists no prior information by the client or any other entity which objects to prefetch. Instead, the prefetching engine is solely responsible to predict future accesses and come up with candidate objects to bring as prefetched objects based on previous access history patterns of the user. Informed approaches are also referred to as semantic/content based prefetching due to the fact that it is often the case that semantic relation is provided as the hint talked about.

b. In their detailed survey, Vijayan and Jayasudha [62] claim that prefetching algorithms could be classified into three, namely, probability based, clustering based, and weight-function based. They have examined algorithms in each of these classes extensively, which are presented in subsequent sections.

c. Walled et al. [11] also presented an extensive assessment of prefetching schemes in which they have classified the approaches as, Dependency Graph based, Markov Model based, Cost Function based, and Data Mining based. The data mining based approaches were further categorized into association rule based and clustering based. Cost function based approaches were also classified further into four: Prefetch by Popularity (i.e., prefetch a popular document), Prefetch by Lifetime (i.e., prefetch a document with maximum lifetime), Objective Greedy (i.e., prefetch a document to optimize a specific configurable metric), and Balanced Metric (i.e., prefetch a document while balancing between various metrics).

d. Thangaraj and Meentachi [1] presented a good survey of prefetching techniques. This work attempted to put various prefetching approaches into some sort of structure. In the process, the authors classified prefetching attempts into three generic classes, namely, web prefetching, data prefetching and other issues related to prefetching. The web prefetching class is further sub divided into cache prefetching (also known as history based prefetching) and semantic prefetching. Data prefetching is further sub-classified into content and context prefetching.

e. When we consider where the prefetching module is built, the approaches can be classified into three, client side, server end, and at an intermediate proxy server [11]. A number of attempts can be cited for implementation of prefetching algorithms in each of these locations.

History based prefetching schemes are ideal to implement either at the server end or at an intermediately proxy server as these locations are often equipped with access log mechanism which keeps track of requests made by users. Hence, history based prefetching in these cases is a matter of forming a method that tries to analyse the access logs to come up with a model/pattern that best describes future requests based on past access trend.

Moreover, the performance of the history based prefetching engine increases in terms of hit rate if it is implemented at the origin server compared to any other proxy location because of the larger user access information observed at the server. However, this approach has its own drawback. Since the information about which objects to be prefetched is created at the server end and then transferred to clients or to proxy server, which in turn actually perform the prefetch request, it will cause a bandwidth overhead. Since server side prefetching pushes the decision to either the client or the proxy, it is also known as push based prefetching. On parallel note, client and proxy side prefetching approaches are called pull based prefetching.

Semantic prefetching approaches, on the other hand, are not recommended to be implemented at the server end. This is because there exists an extra parsing and processing of content involved in these approaches which is computationally infeasible to perform by a server.

The above stated facts make proxy servers relatively better location for implementing prefetching. However, proxy servers are not perfect either. Proxy servers fail to result in personalized prefetching and they lack means to utilize information specific to a user at the client end as their primary focus is on creating a generalized pattern for a group of users. Similarly, they fail to reflect browsing pattern for a single server by all users. Furthermore, as stated earlier, their performance in terms of hit rate is lower than that of server side history based prefetching.

In general, all of these algorithms are implemented at one location, be it client side, server end or at an intermediate proxy. None of these algorithms provide a platform whereby information sharing between the three locations listed is facilitated. This fact limits any algorithm to rely on the information and resources (e.g., computational power) available at the particular location it is implemented on. For example, the server end knows relatively a lot about the content being provided while the client side is in a better position than the other to profile the user, hence, if these ends were able to collaborate in the prediction process, it can only be projected that some level of improvement can be observed in the accuracy. With this approach, the entire process is not limited to log or client side semantic processing only, rather it opens the door for incorporating application level information such as what kind of application we are working with and application specific relationship between data objects.

## 2.2 Specific Approaches

**Interactive and Link Prefetching Scheme**

Interactive prefetching is the most basic and simplest of all schemes in the prefetching arena. With this algorithm, the prefetching engine retrieves all URLs within the web page currently being viewed in the background [62]. This algorithm can archive a higher hit rate of up to 80%, however, at the cost of bandwidth overhead. Link prefetching tries to improve this by allowing content creators provide hint through special tags to the browsers (i.e., client application) to understand and act up on accordingly, in idle time [63]. However, the applicability of this approach could be controversial and sometimes unrealistic as content providers often are reluctant to embed such information. Furthermore, this mechanism forgets to include users' preference and access behaviour because it purely relies on the information provided by content providers.

**Dependency Graph Based Approaches**

The first set of relatively smarter approaches to prefetching were focused on access history examination to form a URL graph, such as Dependency Graph (DG). This approach constructs a graph data structure called dependency graph (DG) that looks like the one shown in Figure 2-1. In this graph the nodes represent web objects and an arc from one node to another represents that the later was accessed after the earlier web object within a specific time window. Each of these arcs is assigned a weight, interpreted as the probability of that particular transition. DG is a prefetching scheme adapted from the context of file system.

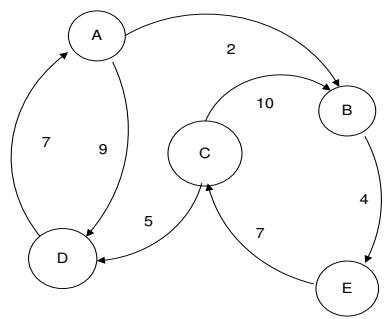

**Figure 2-1:** Sample Dependency Graph

**Markov Model Based Approaches**

An improvement to simple dependency graph algorithm is a Markov Model, which is also popular in the data compression community. The primary task in Markov model approach is to construct a model that could be used for predicting users' next access given session of that particular user.

The input for constructing a Markov model is an array of user sessions considered as training data. Figure 2-2 a) is an example of such a collection of user sessions $WS_1$ to $WS_4$ in which objects $O_1$, $O_2$, $O_3$, $O_4$, and $O_5$ represent possible web objects available for request. From this input a Markov model is constructed that is represented by three components <A, S, T>, where A - stands for set of actions a user could perform (i.e., set of possible objects a user will access), S - for possible states a user could be in (i.e., sessions a user could be in) and T- for Transition Probability Matrix (TPM) which represents the probability of performing an action from a given state, hence an |A| × |S| matrix.

If a single (i.e., currently requested ) object is only considered as the state for predicting future access, the state in the Markov Model will be a collection of single objects, consequently, leading to what is called first order Markov Model (Figure 2-2 b). First order Markov Model is more or less equivalent to simple dependency graph. In first order Markov model, the 'T' value represents the probability a user is requesting an object in 'A' given s/he is at state 'S' which means the user is currently using/requesting the object in 'S'.

On the other hand, if two states/objects (i.e., currently requested and object requested just before this one) are to be considered one ends up with a collection of state that contains a pair of objects which is called second order Markov model (Figure 2-2c). Here, 'T' is interpreted as the probability an object in 'A' is requested given the user has already requested both the objects in S. In general if 'M' objects are to be considered, the model is called an 'M' order Markov Model.

When a large value for 'M' is used, the prediction accuracy increases. In other words, as the number of pages/objects a user accessed resembles more to a session in the training set, the probability that this user will access similar objects in the future from the specified session increases as well. Hence, prefetching can be performed for the object from the session. The problem, however, is even though the accuracy of prediction increases when a larger 'M' is set, the support decreases (i.e., number of patterns observed so far having the specified number of access in sequence). This results in a lower coverage by the patterns discovered.

In contrast, when a lower value for M is set, the accuracy suffers since the model fails to look far enough, same as to the problem of the DG approach. As a result Pitkow and Pirolli [56] proposed an approach called All-K$^{th}$-Order Markov model that started with high order model and if there exists no match, it continues to try lower models until enough coverage is found. A number of literatures proposed a similar approach called Prediction by Partial Matching (PPM) [60], which is known for its adaptive behaviour. PPM is an approach borrowed from the data compression technique [64]. An M-Order PPM keeps Markov models of order j where 1 < j < M using a graph structure. Figure 2-3 illustrates 2$^{nd}$ order PPM.

Markov models assume that a state can be predicted solely based on previous states to some degree of accuracy, which makes it a non-deterministic process. Another direction in user's next web access prediction is using Hidden Markov Models (HMM). HMMs are also popular in natural language processing, weather forecasting, and many other domains that deal with non-deterministic processes.

In its very basic sense, HMM divides states into two, namely, observable and hidden states. For example, if we consider the case of predicting weather condition, in the case of Markov process, prediction of today's weather condition is based on the condition in the previous n days. This is achieved by connection between each of the states, for instance rainy, cloudy, and sunny with a value called transitions probability, which is discussed earlier.

$WS_1 = \{O_4, O_5, O_1, O_2, O_1, O_4\}$
$WS_2 = \{O_1, O_3, O_4, O_2, O_5\}$
$WS_3 = \{O_3, O_2, O_1\}$
$WS_4 = \{O_3, O_4, O_5, O_3, O_2, O_1\}$
$WS_5 = \{O_1, O_2, O_5, O_3\}$
$WS_6 = \{O_3, O_5, O_3, O_1, O_4, O_2\}$

| State | Action | | | | |
|---|---|---|---|---|---|
| | $O_1$ | $O_2$ | $O_3$ | $O_4$ | $O_5$ |
| $\{O_1\}$ | 0 | 2 | 1 | 2 | 0 |
| $\{O_2\}$ | 3 | 0 | 0 | 0 | 2 |
| $\{O_3\}$ | 1 | 2 | 0 | 2 | 1 |
| $\{O_4\}$ | 0 | 1 | 0 | 0 | 2 |
| $\{O_5\}$ | 1 | 0 | 3 | 0 | 0 |

a) Web Sessions

b) 1st Order Model

| State | Action | | | | |
|---|---|---|---|---|---|
| | $O_1$ | $O_2$ | $O_3$ | $O_4$ | $O_5$ |
| $\{O_1, O_2\}$ | 1 | 0 | 0 | 0 | 1 |
| $\{O_1, O_3\}$ | 0 | 0 | 0 | 1 | 0 |
| $\{O_1, O_4\}$ | 0 | 1 | 0 | 0 | 0 |
| $\{O_2, O_1\}$ | 0 | 0 | 0 | 1 | 0 |
| $\{O_2, O_5\}$ | 0 | 0 | 1 | 0 | 0 |
| $\{O_3, O_1\}$ | 0 | 0 | 0 | 1 | 0 |
| $\{O_3, O_2\}$ | 2 | 0 | 0 | 0 | 0 |
| $\{O_3, O_4\}$ | 0 | 1 | 0 | 0 | 1 |
| $\{O_3, O_5\}$ | 0 | 0 | 1 | 0 | 0 |
| $\{O_4, O_5\}$ | 1 | 0 | 1 | 0 | 0 |
| $\{O_5, O_1\}$ | 0 | 1 | 0 | 0 | 0 |
| $\{O_5, O_3\}$ | 1 | 1 | 0 | 0 | 0 |

c) 2nd Order Model
**Figure 2-2:** A sample Markov model

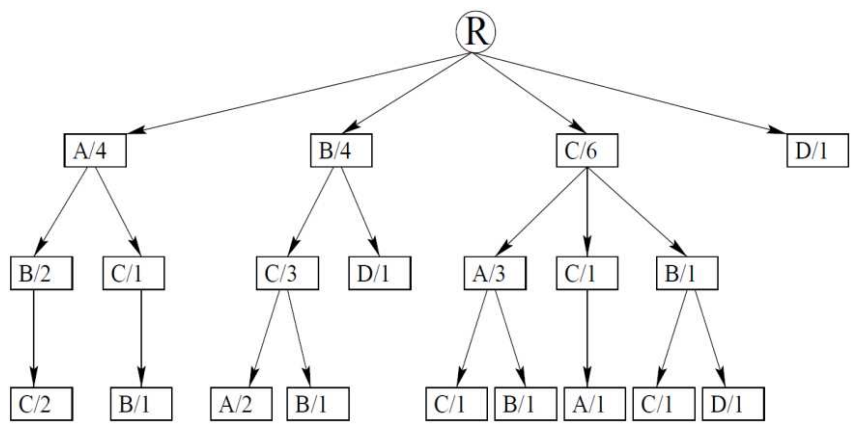

**Figure 2-3:** A sample 2nd order PPM

One observation here is that the transition probability is not time variant. HMM, on the other hand, may consider soil moisture level, wind, and air pressure. For this example, the hidden state of our HMM would be the weather while air pressure, wind, and soil moisture become the observable states.

Association rule Mining has been also used for predicting user's future access, which in its very basic sense means finding a set of if...then rules from a training set of transactions. For example, let I = {$I_1$, $I_2$, $I_3$, $I_4$ ... $I_n$} be a set of n items forming a database D of transaction and T be such a transaction containing a set of items whereby T⊆I. The process of association rule mining is an attempt to find implications of the form A => C whereby A, C⊆I. 'A' and 'C' are called antecedent and consequent, respectively. This is interpreted as "if A happens C will happen".

There exist two important measures in association rules, namely, support(S) and confidence (C). Support refers to the ratio of transactions that contain A and C together out of the entire database D. If, for instance, we have a 2.4% support, this means out of the whole database of transactions 2.4% of them contain A and C together. Confidence, on the other hand, refers to the percentage of transactions that contain C out of transactions that contain A. If, say, a rule has 96% confidence this means that out of all transaction that contain A 96 percent of them also contain C. Put in other words, support measures the relevance of the rule, while confidence measures its certainty.

Zipf's estimator has also been employed in future access prediction. This estimator is based on Zipf's distribution (adopted from the field of information theory) which states that the rank of an object is inversely proportional to its frequency. For instance, in natural language processing words like "and", "is", "a", and "the" are frequent in a corpus while they are poorly ranked in terms of information value they possess.

In the course of semantic prefetching a number of approaches constructed their model by extracting the keywords in anchor text of hyperlinks embedded in web pages. Different attempts employed different learning algorithms for this construction process including Artificial Neural Networks and Naïve-Bayes classifier.

Artificial Neural Network (ANN) is a model inspired by real life neurons, therefore, it possesses the components the later contains to some degree. These are, Dendrites, Synapses and Axon as presented in Figure 2-4 (Source: users.tamuk.edu): A neuron accepts input via its synapses and processes it in its dendrite. Axon is used to connect multiple neurons together.

In ANNs the dendrite, with what we call activation function, and the axons could be associated with a specific weight. The activation function adds the inputs all together after multiplying each by the associated weight. The output axon will produce some output based on this summed value. Normally there is a threshold function that decides the output from this neuron. E.g., if the sum is above X produce m and if it is less than X produce n.  ANNs are formed from connecting these neurons together, because, it is observed that this connection will create a much more powerful model than simple neurons. As it turns out, the network formed is much more powerful than the addition of individual performance. A sample artificial neuron is presented in Figure 2-5 (Source: www.electroyou.it); such a network with a single neuron is also known as perceptron.

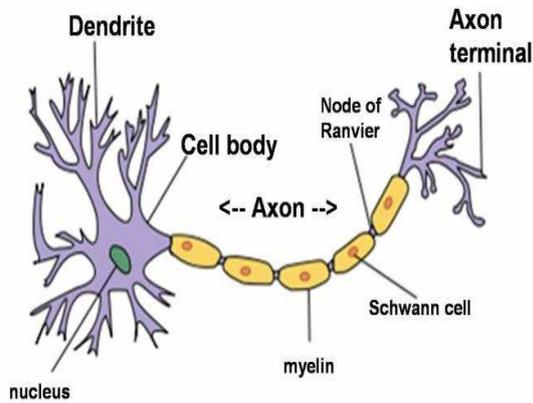
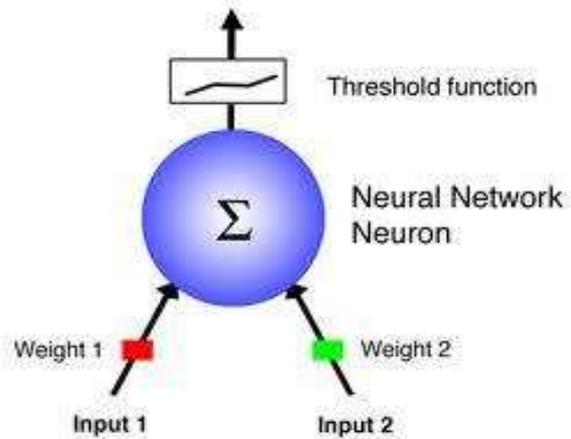

**Figure 2-4:** A sample biological neuron    **Figure 2-5:** A sample artificial neuron

Naïve-Bayes classifier is a supervised learning method based on Bayes theorem. It is particularly suited when the dimensionality of the input is high, that is, when there are sufficiently large numbers of attributes to base our decisions on. For example, if the task is predicting whether or not a customer will buy a computer given his age, income, academic background, Naïve Bayes is assumed to perform fairly good.

The mathematical representation of Naïve Bayes algorithm is given in Equation (1) [65], where p(h) is the probability of hypothesis h becoming true, p(h/D) the probability of the hypothesis h becoming true given the prior knowledge D, p(D/h) refers to probability of D given h and p(D) refers to the probability of having the prior information D in the training data.

$$p(h/D) = \frac{p(D/h)*p(h)}{p(D)} \qquad (1)$$

For instance, in the above example, p(h/D) would be the probability of a customer having the D properties buying a computer (e.g., D can be age = 30-40 and income = 5000-6000 Birr/month, academic background = " College degree ") , p(h) could be the probability of any customer buying a computer regardless of age (obtained from the training data), income and/or academic background, p(D/h) refers to the probability that a customer has the properties D given he has bought computer (obtained from the training data), and p(D) refers to the probability of a customer having the property D regardless of whether he bought a computer or not (obtained from the training data).

As it could be seen from the above example, the probability of a hypothesis h becoming true can be learnt purely from the information presented in the training data. What is more, Naïve Bayes algorithm could learn a very efficient model from small data set.

## 2.3 Performance Evaluation

In the domain of prefetching a number of methods were devised to measure the performance of the prefetching scheme. This section discuss about these evaluation criteria used in other works and the ones adopted for this research.

**Hit Ratio**

Hit Ratio (HR) is the rate of catch hit to the total requests made as the number of cache hit divided by the total number of objects in cache [66]. The mathematical representation of this measure is given in Equation (2).

$$Hit\ Ratio = \frac{\#Cache\ Hit}{\#Request} \quad (2)$$

**Waste Ratio**

Waste Ratio (WR), Equation (3), represents the number of objects in cache that were never accessed compared to the total number of objects [66].

$$Waste\ Ratio = \frac{\#objects\ cached\ but\ never\ accessed}{\#objects\ cached} \quad (3)$$

**Byte Hit Ratio**

Byte Hit Ratio (BHR) is a measurement that does tell about bandwidth saved due to cache hit. It is the ratio of cache hit to total request just like hit ratio but instead of counting requests, this measure is based on the number of bytes transferred (given in Equation 4). Cache hits for large objects contribute more to the byte hit ratio than do small objects. The byte hit ratio measures how much bandwidth the caching system has saved, but there are different ways to calculate it [67].

$$Byte\ Hit\ Ratio = \frac{size\ of\ cache\ Hit\ objects}{size\ of\ all\ request} \quad (4)$$

**Byte Waste Ratio**

Byte Waste Ratio is similar to waste ratio but like byte hit ratio it is measured in terms of the size of the objects rather than merely the count (Equation 5). It could be used to measure the bandwidth wasted due to the prefetching scheme employed. A scheme that aggressively prefetches big objects will perform poorly in this measure.

$$Byte\ Hit\ Ratio = \frac{size\ of\ objects\ cached\ but\ never\ accessed}{size\ of\ objects\ cached} \quad (5)$$

**Effectiveness**

According to businessdictionary.com [68], effectiveness is defined as the degree to which objectives are achieved and the extent to which targeted problems are solved. Effectiveness in the domain of prefetching could be measured by any one of the two hit measures (i.e., hit ratio or byte hit ratio) or by any other measure that gauges the amount of work done.

**Efficiency**

Efficiently is the comparison of what is actually produced or performed with what can be achieved with the same consumption of resources (money, time, labour, etc.) [69]. It is an important factor in determination of productivity. In contrast to efficiency, effectiveness is determined without reference to costs and, whereas efficiency means "doing the thing right," effectiveness means "doing the right thing."

From the above given measures for this work we will consider waste ratio and/or byte waste ratio as measures of cost while hit ratio and/or byte hit ratio gauge the actual work done. Hence, for this work, we have derived Equation (6) as a measure of efficiency and Equation (7) as a measure of byte efficiency. From these equation it can be observed aggressive schemes (i.e., schemes that prefetch too much web objects) will perform poorly in relation with these criteria even if they perform well in terms of hit ratio. Hence, for any prefetching scheme to be applicable, especially in a resource constrained environment such as mobile devices, it should do well in terms of efficiency in addition to effectiveness.

$$Efficiency = \frac{Hit\ Ratio}{Waste\ Ratio} \quad (6)$$

$$Byte\ Efficiency = \frac{Byte\ Hit\ Ratio}{Byte\ Waste\ Ratio} \quad (7)$$

# Chapter 3 - Related work

## 3.1 WebCompanion

Klemm designed a prefetching scheme that estimates the round trip for an object and subsequently retrieved objects having the highest value for this estimation [70]. This method parses HTML documents to list down the embedded hyperlinks using its parser component. Following this process objects resulting with high value for round trip estimation are fetched and cached to reduce user perceived latency. WebCompanion is much more selective than simple interactive link prefetching scheme described earlier.

## 3.2 Cost Function Based Approaches

A number of algorithms attempted to make use of some static measure such as popularity or object lifetime to come up with the objects to prefetch. More specifically, Walled et al. [11], state four measurement criteria for such an approach;

- Prefetch by popularity: retrieve objects that are popular in terms of user requests to a local or proxy cache.
    - **Top 10 Approach**: A scheme that prefetched the ten most popular objects in terms of user request was proposed in [71]. Even though this algorithm is easy to implement, it fails to consider the characteristics of a specific user currently making the request.
    - **Domain Top Approach**: This scheme is an attempt to improve the Top 10 approach by considering popularity of web servers in addition to popularity of web documents as proposed in [72]. This way the popular documents from frequently accessed web servers are registered at a proxy server to be prefetched.
- **Prefetch by lifetime**: This prefetching approach selects a specified number of objects to bring to cache based up on the life time of the objects (i.e., objects having a longer lifetime will be pre-retrieved) [73]. In this scheme, an object that is rarely accessed may be retrieved to cache just because it ranks higher in the lifetime scale.
- **Balanced metric**: Prefetching approaches in this category attempt to balance between lifetime of an object and its popularity. Examples of such algorithms are APL [73] and Good-Fetch [74] (as referred in [11]).
- **Objective Greedy**: These class of prefetching algorithms strive to maximize the performance of the prefetching system in the direction of a certain metric each defines and attempts to uphold (e.g., server load, user perceived latency).

## 3.3 Dependency Graph Based Approaches

Padmanabhan and Mogul used Dependency Graph (DG) algorithm for prefetching [48]. Bestavros [75] also presented a similar approach that uses DG as the underlying data structure but make the prediction decision based on the transitive closer of this graph. However, the result, due to this improvement, was not encouraging enough. Another approach in this stream was presented in [23], which created Double Dependency Graph (DDG) with the intension of differentiating relationships between pages of same server and relationships across web servers.

Even though the user perceived latency could be decreased using dependency graphs, it is at the cost of increased network traffic. Furthermore, the fact that only first order dependency (i.e., dependency between only two pages) is exploited, the performance could be limited [48].

As a result, many of the earlier attempts in prefetching directed their focus on Markov model, which was adopted from prefetching in the context of CUP to memory latency hiding. Padmanabhan and Mogul [48] also used N-hop Markov models in an attempt to improve the performance of prefetching strategies in web caches. Markov models have also been used for classifying browsing sessions into different categories/clusters in [76], which could be one approach towards prefetching. Sarukkai [77] has also used Markov models to predict user's next web page access.

As discussed in the previous Chapter Markov models could be of different order and the higher the order gets the accurate the predictions are. However, higher order Markov models and All-$K^{th}$ – Order models exhibit a high memory requirement due to the large state space associated with it.

Deshpande and Karypis [78] attempted to tackle one of these hitches associated Markov Model based approaches, i.e., the elevated memory requirement due to exponential growth in state space when a higher order Markov Model is used. The Selective Markov Model presented in [78] employees various pruning techniques to reduce the state space requirement of the model constructed. The authors performed pruning based on the criteria, namely, support pruning, confidence pruning, and error pruning. Each of these processes leads to minimized state space Markov Models called Support Pruned Markov Model (SPMM), Confidence Pruned Markov Model (CPMM), and Error Pruned Markov Model (EPMM), respectively.

After the Markov model was constructed then follows pruning, which is the removal of portion of the rules. It is based on the observation that states having low support in the training phase tend to result in lower prediction accuracy. Hence, such states are pruned from the model. In other words, states with minimum frequency in the training set are removed from the model generated. The exact value to start pruning is controlled by a threshold variable $\phi$ known as frequency threshold, which is actually one of the parameters of the model. What this means is that, the SPMM removes any state from the model if the number of training examples supporting it is lower than $\phi$.

The same frequency threshold $\phi$ is used regardless of the order used, which obviously results in reduction of higher order models more often than that of lower order models since the former already have a lower support as discussed earlier. This is assumed to reduce the memory footprint in a greater manner. Furthermore, it can be observed that an actual frequency count is used rather than relative value for $\phi$. This is because trust-worthiness of probabilities is dependent on the actual count and the number of training instances varies depending on the order of the Markov model learned.

However, considering only support value for pruning is not a good idea since the scheme should also consider the probability distribution (i.e., confidence of the action) of actions in the Markov model constructed. What this means is, for instance, if for a state there exist two outgoing actions (i.e., probable actions) and the scheme only considers support count as a pruning measure, the one having the lower value for frequency will be pruned even if it is more probable than the other action. The authors of [78] introduced CMM for such cases. CMM uses statistical methods to determine the significance of the action in terms of accuracy before removing it. In short, if it is significant it will not be pruned even if the support is lower than its counterpart.

The third approach proposed in this work was EPMM in which a validation step is used to estimate the error of each state to come with the pruning decision. In the previous approaches, support and confidence were used to determine error. In EPMM, on the other hand, a validation set is employed as a means to estimate errors associated with a state and an action. The authors adopted two strategies in the construction of EPMM, one that works in a top down fashion (i.e., first validates higher order Markov models and works down to subsets of each) and one that works in a bottom up manner (i.e., first validates first order Markov models and works towards their supersets).

The authors of this work performed their experiment on a collection of real world datasets from two e-commerce sites, a log file of editing commands from Microsoft Word, and a telephone switch alert log. Other than the first two datasets (i.e., those from the e- ecommerce), the datasets are barely related to prefetching. The Authors used such kind of data sets to show the superiority of their work over the All-$k^{th}$-Order Markov Model scheme in terms of prediction, even, for domains other than web prefetching. Furthermore, the pre-processing stage of the aforesaid evaluation ignored image files in the session construction process from the web log datasets.

However, this work did not show the instruction of these pruning criteria, nor it did combine any of them in an attempt to improve the performance of pruning. Furthermore, the authors did not examine the effect of application domain knowledge, which could be represented in some sort of semantic relationship model such as ontology, in improving the prediction accuracy.

With the aim of addressing the aforesaid limitation Mabroukeh and Ezeife [79] presented a Markov model that integrates semantic information represented in a form of domain ontology that is becoming commonplace as more Internet businesses are including them in their online applications. Moreover, numerous services directed toward ontology continue to come into being, for instance, a number of ontology search engines (e.g., OBO Foundry [80], Swoogle [81]) and ontology libraries (e.g., DAML Ontology Library [82], Protege Ontology Library [83], ODP [84], and COLORE [85]) are available nowadays.

Subsequently, the authors of [79] focused on utilizing such kind of information in addition to the Markov models constructed in previous attempts so as to create a model of closer performance to that of All-$K^{th}$-Markov model while keeping a minimum memory footprint. They undertook this task in two directions i) utilizing the semantic information in the process of predicting next access, and ii) employing the semantic information in the process of pruning the All-$k^{th}$-Markov Model.

Before going about in any of the two directions stated, the authors converted the ontology, normally represented in a graph data structure, to a semantic distance matrix M of size N x N, where N stands for the number of web objects/pages involved. An entry in the semantic distance matrix, say $M_{p_i,p_j}$, means there exists that number of edges separating $p_j$ and $p_i$ in the ontology graph. This is, of course, assuming a web object represents one concept only. A sample mapping of pages to concepts is given in Table 3-1 and the Semantic Distance Matrix for these pages is shown in Figure 3-1. Both of these samples are taken from [79].

**Table 3-1:** sample domain knowledge contained in each accessed page

| Accessed Page ID | Actual web page | Ontology mapping |
|---|---|---|
| $P_1$ | /cameras.html | Cameras |
| $P_2$ | /cameras/canon.html | Still Cameras |
| $P_3$ | /chem/fsoln.html | Film developing solution |
| $P_4$ | /film/videofilm.html | Video Film |
| $P_5$ | /elect/dbatteries.html | Dry Battery |

$$M = \begin{bmatrix} & p1 & p2 & p3 & p4 & p5 \\ p1 & 0 & 1 & 5 & 1 & 3 \\ p2 & 1 & 0 & 2 & 2 & 3 \\ p3 & 5 & 2 & 0 & 4 & 8 \\ p4 & 1 & 2 & 4 & 0 & 8 \\ p5 & 3 & 3 & 8 & 8 & 0 \end{bmatrix}$$

**Figure 3-1:** A sample Semantic distance matrix

The first approach proposed, once such a matrix is constructed, computes a weight matrix as given in Equation (8). The weight matrix is a direct combination of the Semantic distance matrix M and the Transition Probability Matrix (TPM). Before the addition is performed, the semantic distance matrix M needs to be normalized because the entries in TPM are all in the range of zero to one. This is done by dividing each value of M by the row sum. Furthermore, TPM represents probability while M represents distance, which means, an increase in TPM adds to the likelihood of being predicted while an increase in M is inversely proportional to this possibility. Hence, in order to bring these measures to the same page, each non-zero entry in the normalized M is deducted from one. Finally, the two matrixes are added to result in the weight matrix stated above. This new matrix is used as the basis for prediction instead of the TPM.

The second approach sets a maximum semantic distance measure η which is used as a fourth method of pruning in Selective Markov Model. In this case, an All-$K^{th}$-Order Model is constructed and then, to reduce the state space and remove contradicting predictions, any branch exhibiting a semantic distance measure lower than η is discarded from the model. This combined with the pruning criteria proposed in [78], will reduce the memory footprint in a great manner.

$$W_{p_i,p_j} = P_{s_i,s_j} + \begin{cases} 1 - \dfrac{M_{p_i,p_j}}{\sum_{K=1}^{n} M_{p_i p_k}} &, M_{p_i,p_j} > 0 \\ 0 &, M_{p_i,p_j} = 0 \end{cases} \qquad (8)$$

In addition, an improvement to All-K[th]-Order Model, more specifically attributed to [79] is the ability to handle contradicting predictions. This work soothes the problems associated with having two or more branches for a given state with equal probability to one another within a given Markov Model, what is known as contradicting prediction. By employing semantic information, it has been observed that it is possible to eliminate such confusions.

This work, however, focuses on integrating semantic information into a specific approach, namely, All-K[th]-Order Model. Hence, no attempt was made to make the approach more flexible and enable integration of semantic information with other kinds of algorithms. Furthermore, the algorithm expects a well-designed and full ontology to be predefined by the content providers end. Even though a rise in number of web applications with this scenario is observed, it is not an all-covering solution. In other words, this approach fails to handle cases where no or incomplete semantic information is available.

Jin and Xu [86] used Hidden Markov Model (HMM) to generate a model that can predict the users' future access. This work is based on the notion that the user access sequence could be understood as a deep search for a concept. With that in mind, the authors constructed a HMM whereby the possible concepts a user is looking for is considered as the observables and the web pages as states.

The first step in the construction of the model, following a pre-processing stage, was the extraction of the required concepts from anchor texts (i.e., labels of hyperlink anchors). This is based on the observation made by the authors that anchor texts are extremely summarized representation of the content of the page they refer to.

With this in mind, the authors constructed a pseudo document, denoted by HyperDoc, using the collection of text strings extracted from the anchor labels. The pseudo document then goes through a number of natural language processing stages including stop word removal and stemming to make it more focused towards the semantics.

Subsequently, all pages located in the web server are investigated to understand the concept they possess. Furthermore, the possible list of concepts a user is looking for is also predicted through the user's access path analysis. All of the pages cited by the page a user is viewing currently are sorted by the degree of their possession of the top n concepts that the user is likely looking for. Out of these pages, the top r pages are considered eligible for prefetching, hence, are retrieved in the back. In their work, the authors found the reasonable number of concepts (i.e., value of n) to consider to be seven. Considering any number higher or lower proved to negatively affect performance of the scheme.

The authors used two performance measures, namely, hit ratio, and session hit ratio (i.e., the rate of accurately prefetched pages from those requested within a given user session). Furthermore, the authors considered a Markov model based scheme as used in [77] for the comparison purpose. What is more, the dataset used was from Labtest web server that mainly focuses on Chinese information processing. Following the evaluation, the work in [86] outperformed its counterpart under consideration in terms of the measures stated above.

Even though the performance evaluation shows superior results, this work is limited in a number of directions. To begin with, the authors did not put into consideration the fact that images and other big objects (e.g., flash files and applets) are commonplace in web technology. The authors, in this work, simply ignored such web objects. In this regard, they failed to show the performance of their scheme using measures such as object hit ratio, which would possibly have reflected the consequence of their decision to ignore such kind of files.

In addition, the authors only considered hyperlink relationships through an anchor tag within an HTML page. However, web pages could be linked with each other through client side scripts such as JavaScript and VBScript. This is common when we consider pages that make use of partial page loading via AJAX technology.

On top of these, the proposed scheme needs to be aware of all of the pages available on the web server. This closely ties the scheme to a server side implementation only, which makes the scheme suffer from all the limitation a server side scheme experiences with no possibility of porting it to any other location.

Another direction towards prefetching is via association rule mining. The authors of [16] claimed to have used an association pattern mining technique that considers all specific characteristics of web-user navigation in the process of prediction of user's future access. One attribute of web resources capitalized on by the authors is its tendency to change frequently in contrast to many other applications such as network file systems.

The aforementioned work utilizes a user's session determination technique that is similar to many other works. The focus of this work was in creating a novel association rule mining technique which considered three factors: a) the order of dependencies between page accesses, b) the arrangement accesses with in a sequence, and c) the noise present in user access sequences due to random accesses of a user which deviates from observed patters.

As a result, the authors in [16] crafted a modified association rule mining technique, more specifically, the Apriori algorithm in the mining process. The modified approach considers sequence of object accesses in addition to simple concurrence that is used by the native Apriori algorithm.

Three performance measures were used on both real and synthetic data. These are usefulness (also known as recall or coverage), accuracy (also called precision), and additional network traffic incurred due to the introduction of the prefetching scheme. Furthermore, the scheme was gauged against a number of approaches including Dependency Graph, Prediction-by-Partial

Matching, and a plain association rule mining algorithm. Consequently, the experimental results show that this scheme outperforms the stated approaches.

Zipf's estimator was used in [87], alongside association rule, to predict next page access. The process starts with a pre-processing of the raw log files, followed by user's session determination (i.e., grouping of request belonging to the same user session together). The authors handled this phase as a clustering task that tries to group user requests using rough set first algorithm by Pawlak et al. [88].

Afterwards, Markov model construction and association rule mining were performed in an attempt to discover patterns within the user web object access sequence. Based on the patterns discovered, when a user requests a web object, the Zipf estimator is used to calculate the probability of accessing the next page given the current request on the basis of all the patterns this object is observed in. The authors, however, did not present any evaluation of performance on their work.

Zhang et al. [89] implemented a predictor module, which basically performs client side prefetching. What makes this work different from other client side prefetching attempts is twofold; a) it is an actual implementation that was integrated into a working product - Firefox, and b) Even though it gives priority to history based approach, this work also considers content based prefetching to come up with the list of URLs to fetch.

With their design, they aimed at developing a configurable, portable and efficient module. Their focus on portability enables an easy repositioning of this module to any location in the caching hierarchy. In other words, this module is claimed to work (following little modification) with other applications such as Squid, which is a popular HTTP proxy.

The aforementioned predictor module first tries to come up with a list of candidate URLs from a history based sub-module that is built by encompassing a PPM model as its major component. If this sub module is unable to come up with five URLs, the content based prefetcher is consulted to fill the list. The number five, in this case, is a predefined threshold value that can be changed by developers employing this module, hence, the term configurable. The content based prefetcher, on the other hand, tries to build a model by analysing the content of the current page and recently viewed ones. When a new page is loaded, the words in the hypertext anchors are compared with the model to predict potential accesses.

Although they tried to combine content based/semantic prefetching approach with that of history based algorithms, it possesses its own share of limitations. Mainly, the work is more focused on implementation issues instead of developing a new algorithm. Hence, it has a number of limitations one being the fact that priority is always given to history based approach regardless of the context. What this means it that the content based approach is always considered as a fall-back rather than being an equivalent candidate to its history based counterpart. This will harm the performance of the system when, for example, we have pages that are highly rich in semantic content.

Another limitation is that it is implemented only at the client end. Consequently, the approach can only make use of information available at this location only, which could be a great hurdle for the algorithm not to perform at a much higher rate than that of its simple client side prefetching equivalents.

A coordinated proxy-server prefetching was proposed in [53]. The authors moved the primary prefetching decision to the proxy and enabled a means to request assistance from the server whenever the proxy is in lack of satisfactory background information for predicting future access. In addition, with this approach, the proxy is able to share any background information it possesses with the server whenever the object is not available in the proxy cache to further improve accuracy of prediction made by the server.

The primary objective of this work was to minimize network traffic associated with a prefetching scheme employed solely at the server end, which compared to implementing a scheme at a proxy server, is high, while keeping the high performance in terms of hit rate normally observed for algorithms implemented at the server end. In other words, this work aimed at improving the accuracy of proxy based prefetching to a value close to that of server based designs while maintaining a minimum network bandwidth overhead which is a common place in server initiated prefetching.

According to a trace-driven simulation performed by the researchers, it has been observed that the performance of purely proxy-based prefetching can be improved notably with a measure close to that of server based prefetching while keeping network bandwidth overhead due to server-to-proxy and server-to-client prefetch pushes to a minimum.

This work, however, only focuses on coordinating proxy and server prefetching which consults client side prefetching engines in an inadequate manner. Furthermore, it focuses merely on moving the PPM prefetcher to the proxy in order to minimize network overhead while keeping the performance close to what it would have been if the prefetching was performed at the server end. It does not attempt to improve the overall performance of the prefetching system by employing different prefetching algorithms at each end that works best for that location and combining these algorithms to improve the performance. These algorithms, if employed, possibly will enrich the overall performance as each could exploit unique characteristics available at each end.

Xu and Ibrahim [90] presented a keyword based semantic prefetching scheme for Internet news service that capitalized on semantic locality rather than temporal locality, which has been the focus of many approaches. This work was a result of the observation that, if a user reads news about a given topic s/he will likely request news content of a similar/related content in the near future.

For instance, a user requesting for news about football in relation to Brazil will probably read about the world cup if it is a current affair. In this case, Brazil and World cup could be considered to be keywords. From this example, it is obvious that the interestingness of a key word fails over time which is a concept dealt with by the authors via an adaptive learning algorithm that changes the significance of keywords.

Accordingly, the authors developed a prototype agent which consists of a perceptron (i.e., an Artificial Neural Network with a single neuron). This perceptron accepts keywords extracted from a URL anchor text with associated weight, which represent the significance of the keyword, as an input and the output is either zero (do not prefetch) or one (prefetch). The perceptron outputs one if the sum of weights of each of the keywords is above a certain threshold value T, and zero otherwise.

The authors used Least Recently Used (LRU) approach to purge web objects from cache. With this approach web objects not accessed for a while will be removed when there is a cache storage scarcity. What is more, the weight of keywords that lead to these web objects is adjusted accordingly, so as to purge them the significant keywords list if their respective weight fails below a certain threshold value.

When a user visits a web page all the anchor elements within the page are extracted and grouped, subsequently the first group of anchors is considered. This is because, the authors stated, the average time a user stays reading a page before advancing to the next link is about 30 seconds. Therefore, the prefetching should be performed before that time elapses, which would be nearly impossible if the scheme considered all of the links in the page. Each URL from the selected group is processed to extract the keywords from its anchor text. The keywords are supplied to the perceptron which comes up with the decision to prefetch or not.

The learning conducted was a supervised one, which means the scheme was informed about its correctness of the prefetching decision it has made. What is more, the learning used Least Mean Square (μ-LMS) - incremental equivalent of gradient decent, making it an incremental learning that adapts to changes in significance of keywords over time. This adaption is controlled by a value called the learning rate. When the learning rate is set high, changes in significance of keywords is felt in the model fast, while if set to a low value, the model will be accurate and more reliable in its prediction.

The model was trained by real time user (one of the authors) that accessed cnn.com looking for two hot news topics at the time of study. Once the model was constructed the evaluation took place on abcnews.com which, at the time, was covering more or less similar news topics.

Hit ratio, byte hit ratio, waste ratio, and waste byte ratio were employed as performance measurement mechanisms. The model performed promisingly in terms of hit ratio and byte hit ration. However, it performed poorly in waste ratio measures, mainly because users will stop looking for similar topic after a while as they are satisfied with the information they have acquired so far. The model, on the other hand, will continue to prefetch the related objects when the user is viewing the last page (i.e., the one just before, s/he loses interest in that topic) which are never requested again.

One major limitation for the proposed scheme is its domain specificity, namely the Internet news service. As a result, the agent should be explicitly turned on and off depending on what the user is currently browsing. The authors acknowledged this and considered adapting this work to other services (e.g., online shopping) as their future work.

A similar work presented in [65] used Naïve Bayes Classifier for predicting future access from a set of keywords (tokens) collected into a repository, same way as in [90]. This did not consider any particular domain at design time. However, the evaluation was conducted only on two scenarios, namely, a prospective student visiting several university websites and a person visiting news portals.

In the direction of Multi-source prefetching Davison [26] examined the possibility of a multisource prefetching scheme that incorporates prediction from multiple algorithms (also known as Sources). In this work Davison considered combining both content based with history based approaches and multiple history based schemes to find a promising level of improvement.

However, when combining content based approaches with that of history based schemes Davison gave priority to the semantic schemes and then considered the history based approaches to fill the space in the prefetch queue that was not filled by the former. This fact makes the approach made by Davison no different than the one presented in [89] in terms of fairness towards all sources.

On the other hand when combining multiple history based approaches no fixed prioritization (i.e., favouring one or more of the algorithms/sources indefinitely) could be observed. Still when combining sources in all cases this work considered a fixed weight for all sources which is inconsiderate of the possible change in their ability to predict future access over time.

## 3.4 Summary

Although more prefetching approaches can be examined in a much more extensive survey thoroughly, more or less, all of them are similar to each other. Existing prefetching schemes are bound to a handful of context information they could leverage from. Most still assume the web to be simple content sharing platform as it used to be in the past.

However, with the introduction of Software As a Service (SAS) and related concepts, the context of today's web exists in a much broader than the one in its infancy. On top of that, the way users access the web has dramatically changed. Users nowadays prefer smaller and ubiquitous devices, which face resource constraints, to connect to the web. What is more, web applications are getting more complex, dynamic and diverse whereby each web application brings its own set of domain specific information that could potentially improve performance of our Internet infrastructure if properly utilized.

Existing prefetching schemes fail to incorporate available context information (e.g., domain knowledge, user location, and so on) in an application independent and generic manner. To give an example, if we consider algorithms in the history based prefetching, they are all focused on predicting future user access solely based on previous user access from history log. It can be observed that the various attempts examined in this Chapter with relation to history based approach either attempt to use a new data mining functionality in the access pattern analysis process (e.g., Markov Process, Association rule, Clustering, etc.) or attempt to modify parameters of a previously introduced web log mining technique in hopes of achieving an improved prediction accuracy or lower cost.

The content based prefetching approaches, on the other hand, are simple anchor text based attempts that barely consult other sources of information while predicting forthcoming user accesses. This, while it seems little, is based on a very important assumption, that is content in the web is solely connected using anchor tags within the HTML pages. This is obviously not the case for contemporary applications that are now using client side script extensively to facilitate and guide navigation.

Furthermore, content based approaches need to be aware of the domain they are working in to make the best out of them, which in turn leads to the need for user intervention for turning them on while there is a need (i.e., when the user is surfing a web in that domain) and disabling them when the user traverses to a domain not covered by the model. Put in other words, all of the approaches fail to incorporate information from multiple sources in their prediction decision (i.e., they are not truly multi-source). What is more, the design of a prefetching scheme is highly tied with the location it will be implemented in. Hence, the prediction could not be moved to a different location on the fly if there is a need based on computational resource availability.

The following listing tries to highlight limitations in each of the prefetching schemes discussed so far according to the different categories they were presented.

**Server end Implementations**

- Even though the server is aware of the available objects more than any location in the system, it fails to capture true user access patterns in the existence of proxy servers.
- Bandwidth overhead since the prefetching decision is made at the server end and transferred to the proxy.
- Even for those users whose access pattern was captured successfully, the behaviour of the client only on that server is captured.
- Semantic prefetching is not feasible as it introduces extra processing cost which makes server end implementation unrealistic.

**Client end Implementations**

- Although semantic prefetching is possible, it is limited to the degree of information that could be used.
- These approaches fail to see what other clients' trend is while predicting future access.
- Attempts are either naïvely aggressive approach or semantic to a very constrained degree.
- Contemporary semantic approaches require user intervention and are domain specific.

**Proxy side Implementations**

- Fail to incorporate application and user specific knowledge that could help in the process, which is available at the server and client end, respectively.
- Fail to capture personalized preference and not ideal for semantic prefetching (even though relatively better than the server end) as the parsing and processing of content takes up huge resources.

**History based approaches**

- New objects that have never been accessed are virtually invisible to such an approach.
- Simply do not make sense in some domains whereby the history of one user cannot help to characterize the future access of another user.
- It can only be projected that their performance will diminish as the web is getting more and more dynamic (i.e., new web objects are created more often than ever) because identifying useful patterns (i.e., those patterns with enough support) in time would become difficult.

**Content based/semantic Implementations**

- Contemporary schemes are completely domain specific making them less transparent than their history based counterparts.
- Even though following the advances in Web 2.0 more and more content providers are making semantic information about their content available in one form or another (e.g., OWL and RDF), such information is not always available for all scenarios as a canon.
- Domain knowledge is more or less absent at the client end.
- Perform poorly compared to current history based approaches.

**Multi-source Prefetching**
- Most attempts in multi-source prefetching were merely limited in integrating handpicked few algorithms that the overall scheme is tightly coupled with.
- Some only use algorithms that are more or less related and could share the underlying input data directly rather than sharing their prediction result which makes it unscalable.
- Because most schemes focused on the individual algorithms used in their respective work rather than the integration's flexibility to accommodate new algorithms the end product was hardly scalable.

Therefore, what is needed is a framework that allows incorporation of predictions from multiple sources. The framework should enable absorption of context without making it domain dependent as the current semantic approaches are. By making the framework capture the current context via a specialized algorithm for it and act accordingly, the need for users' intervention could be eliminated.

Furthermore, such frameworks enable the collaboration of algorithms that are 'expert' in a specific context with others that specialize on different one. The specialized algorithms could be of either history-based or content-based approaches. This specialization by the semantic prefetching algorithms is the result of the amount of human expert knowledge integrated into them while they were designed. On the other hand history based schemes vary in their expertise and specializations based on what type of learning algorithm they have employed or based on where they are implemented and which portion of the log data is available to them (often influenced by where they are implemented).

# Chapter 4 - The Proposed Framework

## 4.1  Overview

The primary goal of this work is to design a framework that enables collaboration of contemporary prefetching algorithms when making the final decision as to which object to prefetch. This collaboration enables the prefetching scheme capture multiple context information and exploit the fact that algorithms vary in their ability to perform prediction under different context. In other words, an algorithm 'a' may perform better than another, say 'b', under a certain context while the reverse is true for dissimilar context. Context here could include where the algorithm is run, which set of information are available to it and so on.

In the process of fulfilling the aforementioned goal, the designed framework opted for an approach that does not fixate on particulars of any existing prefetching algorithm. Instead it focused on common characteristics exposed by all. As stated in previous Chapters, all existing algorithms emphasize on the prediction of objects to prefetch as that is the critical part of any prefetching scheme. Once they itemize the objects to prefetch they will go on to bring these objects to local cache. How and when they do the actual fetching of these objects is more or less similar even though there might be a slight variation from one scheme to another. The way each algorithm generates the list, however, has a principal difference.

To avoid being caught in the inner working of any algorithm the proposed framework only considers the shared characteristics of these schemes, that is, list of objects to be prefetched as predicted by the algorithms. As stated above, schemes vary in how they make the prediction and what source of information they use to do so. However, they all end up in creating a list of objects to prefetch. Even though which object is included and how confident the scheme is in predicting that object (i.e., the probability that the object is going to be requested by the client in the near future) may differ for each source, the end result could be perceived to be similar structurally for all schemes in a highly conceptual level.

Therefore, the framework proposed conceptualized prefetching algorithms to be merely components that accept the current context[1] and provide a list of objects to prefetch each accompanied by a confidence value. In the existing algorithms these objects (i.e., those generated by the algorithm in play) will be fetched to cache in descending order of confidence value, while for the framework proposed each of these lists are perceived to be candidate lists and an aggregate list is formulated based on these candidates through methods to be discussed in later sections.

Because of the abstract view of algorithms adopted, it is possible to implement the framework over a distributed system spanning from the client to the content origin server. This is possible since each algorithm could be implemented at any location as long as it is able to generate a list of objects to prefetch which will be shared via any message communication mechanism as a simple list. This way all algorithms could be implemented at the location they best perform on (i.e., client, server or intermediate proxy).

---

[1] Context here will include the current user's navigation history in addition to other information relevant to the prefetching scheme in use (i.e., history based or semantic prefetching) as defined in Chapter 2.

Introducing a new algorithm to the framework will also be simple as long as it is able to produce a list of candidate objects with an associated confidence value. Considering this is the case for virtually all prefetching algorithms, it is a safe bet to say the framework works with all contemporary prefetching algorithms.

We opted for algorithm level integration, among other integration forms such as integration at the data source level and so on, because it allows capturing of varying context and dynamic management of sources[2] in terms of adding and removing sources as they come available.

## 4.2 Functional Model/Architecture

### 4.2.1 Overview

This section presents a general architecture of the new model which is designed to achieve the primary design goal as stated in the previous section. The components in this model could be implemented at any location (i.e., client, proxy and/or source server). However since proxy servers are in a functionally intermediate distance from both client and the server and due to the nature of request flow which is from the client to the proxy then to the server, it can only be projected that the final decision making could perform better (i.e., Aggregator) if implemented here. In addition, since their functions are closely related, it is also ideal if Aggregator, Reputation Manager and Source Repository are implemented at the same location to one another to minimize bandwidth consumption due to extra communication between these components. The functional architecture is presented in Figure 4.1 which is further explained in the subsequent section.

One important assumption we made while designing this framework is that the prediction suggestions from each of the sources will be delivered to the suggestion manager through an alien mechanism, hence, we will not address issues only pertinent to prediction delivery. This assumption is a result of our observation that many works have addressed this subject through techniques such as piggybacking [48, 91].

---

[2] *Source* here refers to a single prediction algorithm within the proposed framework (i.e., contemporary prefetching algorithm) while the data such algorithms use as an input is referred as *data source*.

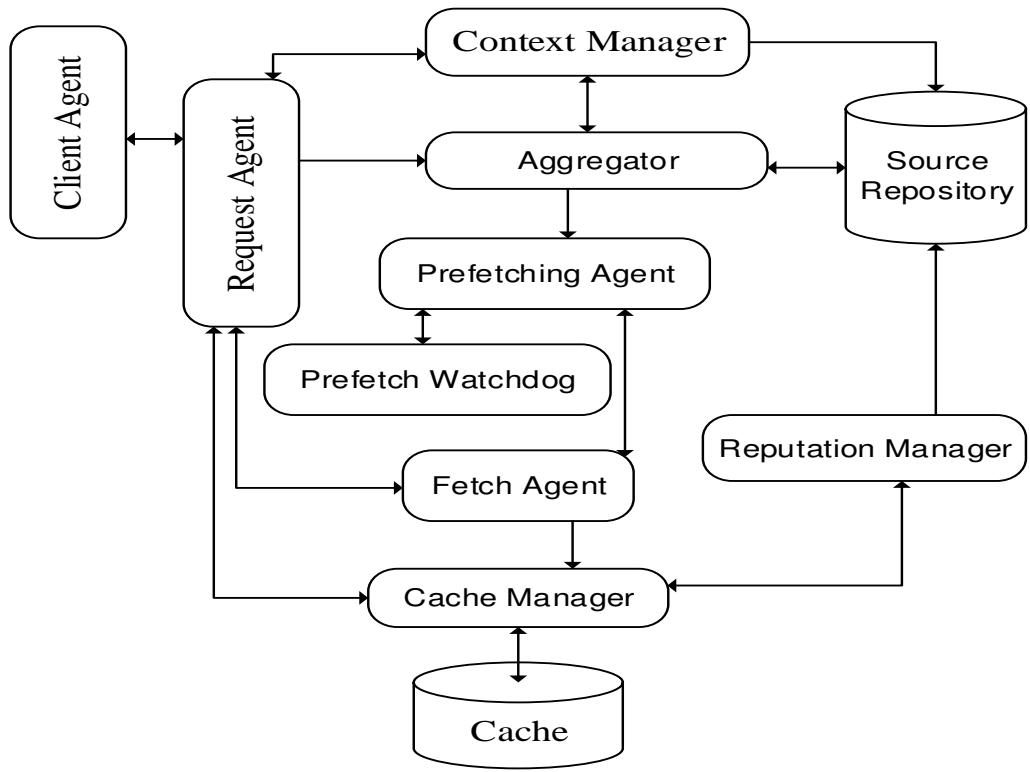

**Figure 4-1:** Functional Architecture

4.2.2  Components

**Request Agent**

The primary reason for having this component is to hide the difference in how requests arrive when the framework is implemented on different locations. For example, if the framework is implemented at the proxy or server end, requests arrive via the network from remote clients hence the framework will return the objects in the same way the request arrives. In such cases, the client's cache is completely invisible to the framework. As a result, the objects fetched have to be sent directly to the client in a predefined means of communication.

On the other hand if the framework is to be implemented at the client end we can assume the cache to be directly accessible by the framework and the client. Having the cache accessible uplifts efficiency of the entire system because only a small signal has to be transmitted to the client stating the object has arrived. When the cache is not accessible the object itself has to be transmitted from the local cache of the framework to the remote cache which is an extra cost especially if the object is large in size. Through this component, the above-mentioned difference could be abstracted from other components of the framework and systems (i.e., client agents) interacting with the framework. The general flow of the requesting process and how new objects are stored to a cache is presented in Figure 4-2 and Figure 4-3, respectively.

**Context Manager**

The reputation of an algorithm is context dependent; hence, a source may have a higher reputation than another source in a given context while the opposite is true for a different context. This engine, therefore, is responsible for enumerating and capturing of context relevant for suggestion analysis (i.e., aggregate list formation from candidates), prefetching and reputation calculation.

One example of context applicable in the prefetching world could be the type of web application the prefetching is done for (e.g., simple static WWW site, dynamic WWW site, and domain specific web application such as enterprise application or SAS, etc.). In this example, a prefetching scheme may be developed for a particular application domain that exploits the domain specific knowledge which is not applicable in other domains.

For instance, if the user is currently navigating web based medical information system a prefetching algorithm that is based on the special relationship of medical documents will have a different list to generate compared to a generic prefetching algorithm that focuses on hyperlink relationship of documents. Similar explanation could be given for other application examples as well, say web based news service where prefetching algorithms that utilize the context already exist [90].

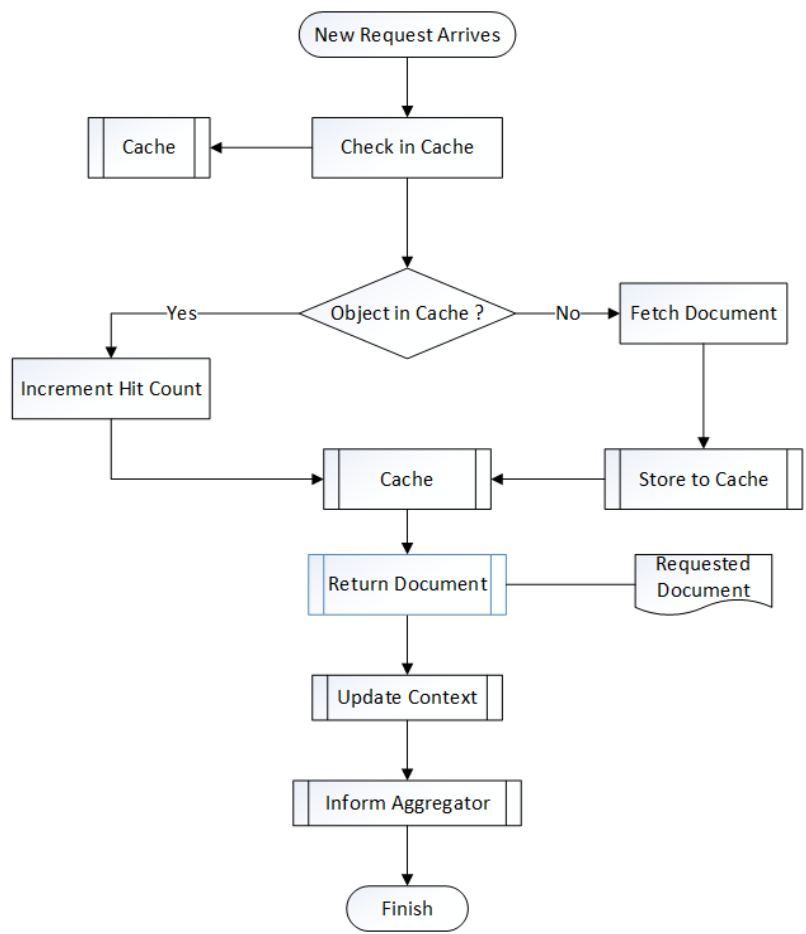

**Figure 4-2:** Flow of new request processing

With the Context Manager component the need for the user to explicitly switch between prefetching schemes could be eliminated with its Context Profiler sub-component. In particular this sub component examines web requests to extract context parameters. These parameters may include, for example, the type of application, spatial and temporal information in relation to the request, and so on. These parameters will be extracted only if there is an algorithm/source that utilizes them for making a prediction of users' future access.

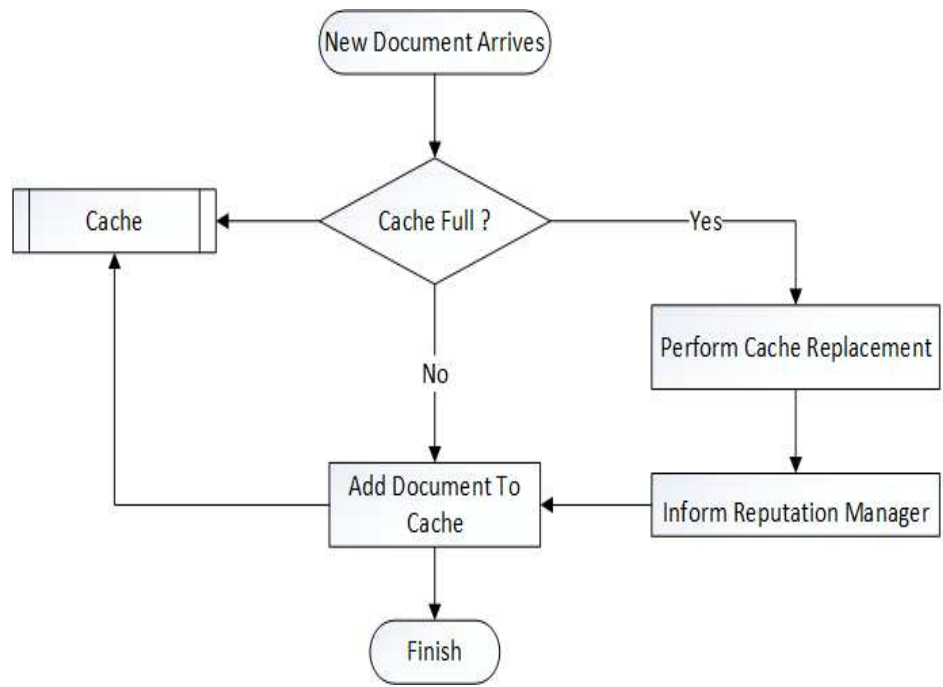

**Figure 4-3:** Flow of storing document to cache

What is more, with this framework no algorithm will need to be technically powered on/off completely rather it will be informed of the current context so that it abstains itself from suggestion if it has no knowledge of the context. If the algorithm has little knowledge of the context it will still contribute to the overall prefetching scheme. However, its degree of awareness in that context will be controlled by the reputation it possesses for the context.

Furthermore, even in cases where the prefetching algorithm is unaware of its specialization in collaboration with reputation management the framework will learn this fact through time. This eliminates the need to drastically modify existing algorithms in order for them to be part of this framework. In other words, the framework will learn the context on which an algorithm performs optimally.

The context profiler is a collection of plug-in routines each extracting a different class of context information. Each source, when registered to source repository, enlists a set of routines (as a plug-in) that are capable of extracting context values that it finds useful to make a prediction in the prefetching process. What the context manager does, therefore, is simply call each of the routines it enlisted iteratively.

**Source Repository**

Source Repository is a component of the model that keeps track of all prefetching suggestion sources registered in the model (i.e., prefetching algorithms). Each source will be providing a list of objects to prefetch under a certain context. The list handed by these algorithms to the model is analysed by the Aggregator to come up with the list of objects to actually bring to cache.

The objective of having this component is to enable dynamic addition and/or removal of algorithms into and out of the framework. When there is a need to include a new algorithm (this is often because it captures a context that is not covered by those before it) it could be done easily without the need to modify the entire framework. This is possible because the proposed framework has a high-level abstract view of involved algorithms.

**Aggregator**

This component of the model will accept prefetch suggestions from all sources registered in the Source Repository. For efficiency reason Sources could be grouped into a set of classes according to the cost associated with obtaining the list. What this means is Sources that are far from the Aggregator are grouped differently than those that are close to this component[3]. This way the Aggregator first consults close by Sources and if it finds enough number of objects to prefetch satisfying a minimum threshold total confidence those that are found far are not consulted to minimize cost.

The actual number of suggestions each source provides is affected by configurable variable of the framework known as prefetch length (p). This value should be selected carefully since a very large number will result in a cache waste as objects with a low probability of being accessed will be brought about. On the other hand a very small value for p will have the same effect as not having a prefetching scheme because it will result in a higher cache miss.

Once all Sources that are found to be necessary to consult are consulted the Aggregator accepts a list of proposed objects to prefetch from each. This consultation and proposal process could be performed in a separate request response transaction or could be embedded with a normal request response flow using piggybacking mechanism as discussed in other works [92].

The list of objects to prefetch that is produced by each of the participating algorithms is considered to be a candidate list and each object in every list will be associated with a confidence value. Confidence refers to the probability of the object to be requested in the near future as calculated by the source responsible for the list given the current context. Which objects are to be included in a list or the confidence value each will be associated with is the decision of each algorithm and no prefetching algorithm is expected to be dependent on the other to perform this task.

---

[3] Distance in this case is measured in terms of cost associated to obtaining the information. A *Source* which requires lesser cost to come up with the candidate list and transfers it to the *Aggregator* is perceived to be close and vice versa.

All objects from all the candidate lists are amassed to a single list by the Aggregator. When producing this aggregate list if an object appears in two or more candidate lists it will be inserted into the aggregate list only once, but will have multiple confidence values (i.e., one for each of the lists it appears in). On the other hand, objects that fail to appear on some of the candidate lists will still be included in the aggregate list, here however, the confidence value from the algorithm that did not propose it is assumed to be zero.

Once the aggregate list is formed, we have crafted Equation (9) to calculate total confidence for each object. This equation is a modification of the technique presented in [26], which is basically a formula to calculate weighted average, that accommodates weight values that vary over time (i.e., reputation in this case) and multiple context. The equation acknowledges the fact that each source will have a different weight/reputation in different context.

In the equation, $C_a^o$ refers to the aggregate confidence for an object O and S is the number of Sources while $R_i^j$ is the reputation of the $i^{th}$ Source under the $j^{th}$ context. $C_i^o$ refers to the confidence Source i exhibited for the object O. The possible number of contexts and the current context among them is identified by the Context Manager component of the model.

Following total confidence calculation the aggregate list is sorted in descending order of total confidence. Then after, elements below the $n^{th}$ element are chopped of the list. Also, objects that already exist in cache are removed from the list. The resulting list is handed over to the prefetching agent.

$$C_a^o = \frac{\left(\sum_i^S R_i^j * C_i^o\right)}{S} \qquad (9)$$

**Prefetching Agent**

Once invoked when the resource is available this agent will request objects from the originating server through the Fetch Agent which will add them to cache. While adding objects to cache, the Prefetching Agent will attach prefetching related information to the object such as the confidence each Source associated with this object in the current context. As stated previously, if a Source did not suggest the object under consideration, its confidence value on the object will be set to zero.

Primarily this component is introduced because normal fetch requests are different from prefetch requests in their nature and intent. For example, under normal circumstances ordinary fetch requests are intended to satisfy client request, hence may be entertained directly from cache without a need to contact the originating server if the object exists there. On the other hand prefetching requests are always about bringing objects to cache from the originating server.

In addition, ordinary requests will have an effect on prefetching context which subsequently affects which objects are to be prefetched while prefetching requests are merely bringing objects to cache without an effect on the aforesaid context. Furthermore, ordinary fetch requests have a higher priority than prefetching requests. The overall flow of prefetching is presented in Figure 4-4.

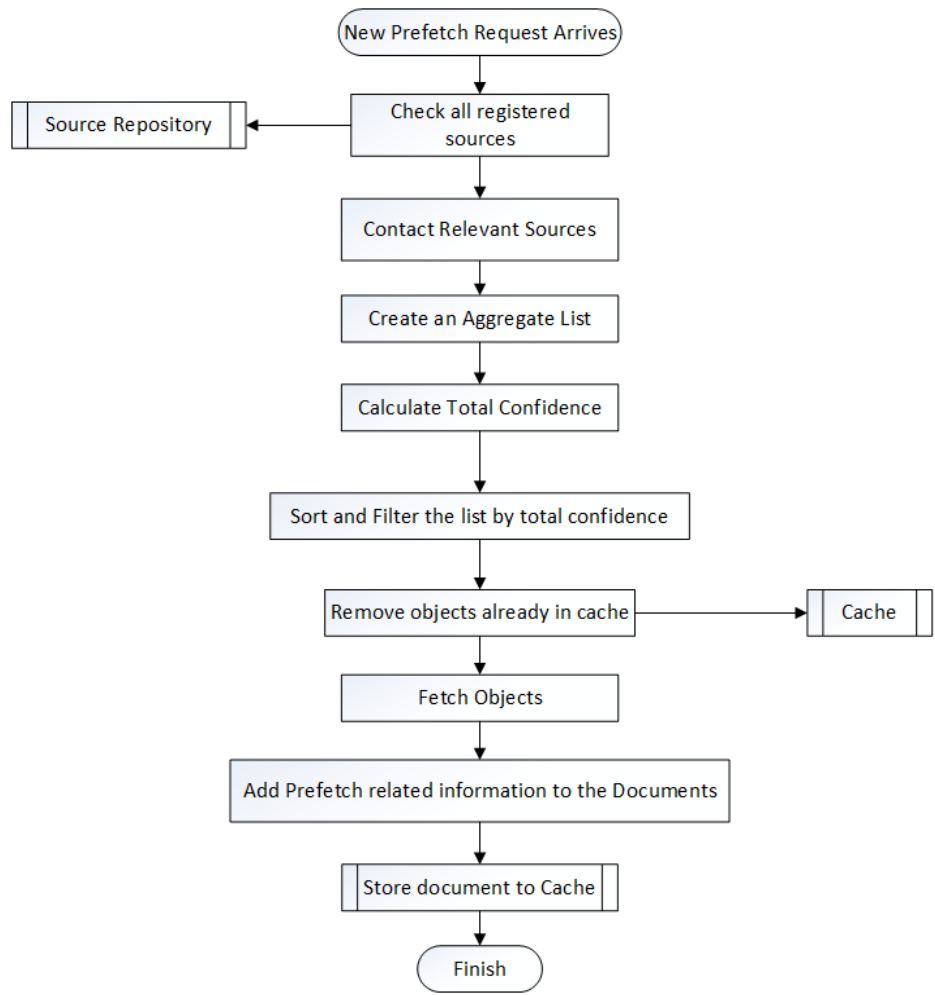

**Figure 4-4:** Prefetching Work Flow

**Fetch Agent**

This component is responsible for accepting fetch requests from either Prefetching Agent or Request Agent. Once the object is brought to cache if the component invoking the service of fetching is the Request Agent it is notified about this happening. If the invoker is the Prefetch Agent it will be handed the object to transform it to a Prefetched Document which is merely a modified version of a simple Cached Document. This transformation is needed to indicate the document is prefetched and associate prefetching only related information with the object.

**Cache Manager**

Cache Manager is responsible for issues related to cache management including cache size decision and replacement policy. In general, this framework is independent of the cache replacement policy adopted; hence, it can work with any cache management policy adopted. A slight modification to the policy is required, however, that makes sure the reputation manager is informed of the cache replacement/removal events.

By default, however, the framework uses List Recently Used (LRU) cache replacement policy as it is the relatively simplest of all to implement with a fairly good performance.

**Reputation Manager**

Every algorithm in the Source Repository is assigned a list of numerical values known as reputation, one for each known context (i.e., identified by the Context Manager). For a given context, an object in the candidate list from a Source with a high reputation will be valued higher than its counterpart from an algorithm with lower reputation for that context, even if both are assigned equal confidence by their respective source. More specifically, the confidence value proposed by a Source will be magnified by the degree of reputation that Source has for the particular context.

This reputation value is updated after every cache removal to adapt a dynamic environment where algorithms are rewarded for their successful predictions, while they are deprived of this reward when they fail to predict a relevant object. When an object is removed from cache based on the cache replacement policy adopted, the reputation updating takes place. If the object was accessed at least once reputation of all algorithm(s) (in the respective context) that suggested it will be rewarded according to the degree of confidence each exhibited on that object and the number of times it was accessed. If the object was never accessed no reward will be given to the algorithm.

In other words, what we need is to increment the reputation of sources in a given context if the object they predicted was useful. Otherwise (i.e., the predicted object was not accessed), the Reputation Manager should not make any increment to this reputation value of the source. In a sense, if all algorithms that did well are rewarded and those that performed poorly are passed over, this lack of reward could be considered as punishment for algorithms that resulted in wastage due to inaccurate prediction.

The degree of reward needs to be affected with both the confidence of the source associated with the suggestion and a learning rate value $\delta$. The learning rate is required as in most learning techniques to control the degree in which rewards update the reputation of sources. Having a big learning rate $\delta$, makes the framework dynamic that rewards sources aggressively for small amount of success. Having a small $\delta$, on the other hand, will slow down the change in reputation.

What this means is if an algorithm, say $S_1$, suggested an object with a certain confidence value, say $C_1$, and another algorithm, say $S_2$, suggested the same object with a confidence value of $C_2$, when that object is about to be removed with hit count of $X > 0$, then, $S_1$ will be rewarded with a higher degree than $S_2$ given $C_1 > C_2$ and vice versa. To reflect these facts, we have derived Equation (10) to control the degree of reward based on the requirements stated. In this equation $R_t^j$ and $R_{t-1}^j$ are the new and old reputation for $j^{th}$ context at time t and t − 1 respectively. H is the hit count of the object and C refers to the confidence exhibited by the Source about the object under consideration. The tan$^{-1}$ is a sigmoid function used here to minimize the effect of a single update to make the reputation change smoother.

If the object to be removed has a cache hit value of zero no further action is taken with regards to reputation management (i.e., no reward is given to any of the participating algorithms). This implies no algorithm will get an increment due to suggesting a wrong object to prefetch.

$$R_t^j = R_{t-1}^j + \delta(tan^{-1}(H * C)) \qquad (10)$$

Having the confidence value involved when updating reputation of a Source is extremely important so as not to affect Source that were not involved in the prefetching of an object. Having the confidence value of an object for a particular Source set to zero when that algorithm did not include it in its list means the algorithm is not participant in the prefetching of this object, hence, will not be affected at all as it can be seen in Equation (11) when C in Equation (10) is set to zero.

$$R_t^j = R_{t-1}^j \qquad (11)$$

**Prefetch watchdog**

Prefetch watchdog is accountable for making sure all prefetching requests are performed on idle times within the system. The rationale for having such a component is to avoid the excess overhead introduced as a result of employing prefetching scheme. With this regard, when to perform prefetching, how many number of objects are to be fetched at a time and the amount of resource allocated for prefetching is controlled by this watchdog.

## 4.3   Example Scenarios

In this section we present examples that we think will further explain how the proposed framework functions. First we illustrate the configuration we assume to have for all of the subsequent examples. Then after, we present various examples that typify how the different components work in selected scenarios.

**Configuration**

For the entire subsequent example we assume to have the framework implemented in the network of a health care provider organization (i.e., a hospital) with all components of the framework running on a proxy server owned by the institute. We assume that the implemented framework incorporates two history based approaches and two content-based algorithms each with a specific domain of specialization.

The first content based algorithm is "expert" on news service as the second one is on medical domain with a hypothetical nationwide medical record system implemented in mind. Developers of these systems are not expected to perform prefetching within their respective application rather the only requirement for them to participate in the framework is to provide an algorithm that exploits the semantic relationship amongst objects of the specific domain to predict a list of possible objects to be accessed in the near future.

Medical Domain is not as dynamic as other domains such as news domain are (i.e., objects do not change frequently) but it is risky for prefetching due to its' big sized documents. By risky we mean, making one wrong decision would result in prefetching of big objects. The news domain, on the other hand, does not contain big objects, per se, but it is extremely dynamic for history based schemes to capture relevant patterns in time. We also assume that the two domain specific semantic algorithms are designed by the content provider to run on the client machine hence are shipped with the content as most current semantic approaches do.

The two history-based schemes, on the other hand, are implemented at the proxy and the originating server. These algorithms are Dependency Graph (DG) based approach and a Prediction-by-Partial Matching (PPM) scheme. The DG scheme works on the proxy log while the PPM works based on the server log.

**New Request for non-cached document**

A user requests an object (web page) from bbc.co.uk for the top news of the day which apparently is about a football match on the English primer league to be held that day. The client agent (i.e., Web browser) sends this request to the framework's Request Agent. This agent checks its cache to see if the object exists and finds out it does not. Hence, it passes this request to the Fetch Agent to retrieve the object from its original location (i.e., bbc.co.uk).

The Request Agent, in parallel, informs the Context Manager, which is responsible for extracting relevant context information from the request detail, about this request and its details (e.g., client profile, time of request and other parameters associated with the request). Once it examined the request to extract context values using the registered routines of all the participating algorithms/sources, it returns this context information to the Request Agent, which informs the Aggregator about this change in context as a suggestion to recalculate its prefetch queue. This re-calculation is meant to assist the queue cope with the change in context and enable the framework adapt to user preference change over time, among other things.

The Fetch Agent, meanwhile, retrieves the requested object to local cache and tags it with information relevant to cache management such as time of caching, context of caching, and reason of retrieval (i.e., ordinary request not prefetch request). Once it finishes its tagging the Fetch Agent informs the Request Agent of this happening which in turn returns the object to the requesting client.

The Aggregator decides to re-calculate the prefetch queue. If there are pending requests the agent will not stop these requests but will do the recalculation to change the remaining part of the queue. To perform this recalculation, the Aggregator queries the Source Repository to get the list of all relevant registered algorithms which it then passes the context information to. We said relevant algorithms because, as stated in a previous section, sources could be grouped based on distance to make the framework more cost effective by only contacting minimum number of required sources/algorithms when some are located far from the aggregator compared to others. In this case, however, we will not consider distance; hence all algorithms are to be contacted.

The list of algorithms together with their reputation value for each of the two contexts (i.e., news service context (context 1) and medical domain context (context 2)) is presented in Table 4-1. Each of these algorithms then returns its top five predications (as per the configuration in which the Prefetch Length is set to five) based on the context information they obtained from the Aggregator. The list of objects each algorithm suggested together with their respective confidence values are given in Table 4-2.

**Table 4-1:** Registered sources and their reputation in the known contexts

| No. | Name/Description | Type | Reputation in context 1 | Reputation in context 2 |
|---|---|---|---|---|
| 1 | News service algorithm | Semantic | 2.0 | 1.0 |
| 2 | Medical record Algorithm | Semantic | 2.4 | 1.3 |
| 3 | Dependency graph based algorithm | Access history | 1.1 | 2.2 |
| 4 | PPM based algorithm | Access history | 3.1 | 1.2 |

**Table 4-2:** List of predictions each algorithm made together with the confidence values

| Algorithm | List of prediction {Object: confidence} |
|---|---|
| 1 | {O3:85.3},{O1:65.0},{O2:60.1},{O4:54.7},{O6:51.5} |
| 2 | {O4:90.6},{O3:72.3},{O5:66.9},{O2:61.8},{O7:56.4} |
| 3 | {O5:87.2},{O4:85.1},{O3:77.4},{O1:69.2},{O6:52.8} |
| 4 | {O3:98.4},{O5:86.7},{O2:81.9},{O4:75.4},{O7:68.6} |

The Aggregator then calculates the amassed list of five objects (given in Table 4-4) using Equation (9) with the list in Table 4-2 (rearranged as in Table 4-3) to serve as an input for Equation (9). It can be seen in Table 4-3 that objects that were not incorporated in the list of an algorithm will have a confidence value of zero for that source. The current context for the request is calculated by the context manager to return a categorical value that represents one of the know contexts (i.e., context 1 and context 2), in this case one.

$$C_a^{o_1} = \frac{(2.0 * 65) + (2.4 * 0) + (1.1 * 69.2) + (3.1 * 0)}{4}$$

$$C_a^{o_1} = 51.53$$

$$C_a^{o_2} = \frac{(2.0 * 60.1) + (2.4 * 61.8) + (1.1 * 0) + (3.1 * 81.9)}{4}$$

$$C_a^{o_2} = 130.60$$

$$C_a^{o_3} = \frac{(2.0 * 85.3) + (2.4 * 72.3) + (1.1 * 77.4) + (3.1 * 98.4)}{4}$$

$$C_a^{o_3} = 183.58$$

$$C_a^{o_4} = \frac{(2.0 * 54.7) + (2.4 * 90.6) + (1.1 * 85.1) + (3.1 * 75.4)}{4}$$

$$C_a^{o_4} = 163.55$$

$$C_a^{o_5} = \frac{(2.0 * 0) + (2.4 * 66.9) + (1.1 * 87.2) + (3.1 * 86.7)}{4}$$

$$C_a^{o_5} = 131.31$$

$$C_a^{o_6} = \frac{(2.0 * 51.5) + (2.4 * 0) + (1.1 * 52.8) + (3.1 * 0)}{4}$$

$$C_a^{o_6} = 40.27$$

$$C_a^{o_7} = \frac{(2.0 * 0) + (2.4 * 56.4) + (1.1 * 0) + (3.1 * 68.6)}{4}$$

$$C_a^{o_6} = 87.0$$

**Table 4-3:** The list of all suggested objects and the confidence exhibited by each algorithm

| Object \ Algorithm | 1 | 2 | 3 | 4 |
|---|---|---|---|---|
| $O_1$ | 65 | 0 | 69.2 | 0 |
| $O_2$ | 60.1 | 61.8 | 0 | 81.9 |
| $O_3$ | 85.3 | 72.3 | 77.4 | 98.4 |
| $O_4$ | 54.7 | 90.6 | 85.1 | 75.4 |
| $O_5$ | 0 | 66.9 | 87.2 | 86.7 |
| $O_6$ | 51.5 | 0 | 52.8 | 0 |
| $O_7$ | 0 | 56.4 | 0 | 68.6 |

**Table 4-4:** List of top five objects in terms of aggregated confidence value

| Object | Total confidence |
|--------|------------------|
| O3     | 183.58           |
| O4     | 163.55           |
| O5     | 131.31           |
| O2     | 130.60           |
| O7     | 87.0             |

The Aggregator then passes this combined list to the Prefetching Agent which will consult the Prefetch Watchdog to start the actual retrieval of objects by calling the Fetch Agent. The Fetch Agent retrieves each of the specified objects to cache and tags it with information relevant to cache management as it did with ordinary request. This time, however, it will inform the Prefetching Agent instead of the Request Agent when the objects arrive. Moreover, the reason for caching is set to prefetching instead of ordinary request as it did previously.

Once the objects are stored in cache and are tagged by the Fetch Agent, the Prefetching Agent will add further tags which are again required for cache management, but this time only applicable for prefetched objects such as the confidence exhibited by each of the participating sources. This addendum information will come handy when we re-calculate the reputation of the algorithms.

**New Request for cached document**

The user, through with the previous web page, requested to see a new web document, $O_1$, which is actually cached by the framework due to a prefetch. The request from the client agent arrives to the framework in a similar fashion as it did in the previous example (i.e., to the Request Agent). But this time, the Request Agent, when it checks the cache, finds out the object already exists. Hence, it updates the cache hit information of the object in cache (i.e., hit count and last hit time) while sending back the object to the client. Updating the hit information is necessary for the Cache Manager and Reputation Manager to work properly.

**Cache replacement**

Due to a shortage in cache space the Cache Manager decided to remove two of the Least Recently Used (LRU) items, say $O_5$ and $O_2$ from the set of objects prefetched in the previous example. And let's also assume $O_5$ was accessed three times and $O_2$ was never accessed since they were brought to cache. Before removing each of these objects, the Cache Manager checks if caching of the object was a result of prefetching or ordinary request. If it is cached due to an ordinary request, it simply removes the object without further action. However, if the object was cached do to prefetching it calls the Reputation Manager and passes the removed object for it to update the reputation of participating sources/algorithms using Equation (10) as given below.

We assume the reputation of each algorithm to be similar to the one in previous example as depicted in Table 4-1. Furthermore, this framework is configured with a learning rate value δ of 0.5. Following this update to the reputation of each source will become as given in Table 4-5. Following removal of $O_5$ reputation of each source in the first context (context 1) is recalculated since item $O_5$ was cached under that context.

$$R_t^1 = 2.0 + 0.5(tan^{-1}(3*0))$$

$$R_t^1 = 2.0$$

$$R_t^2 = 2.4 + 0.5(tan^{-1}(3*66.9))$$

$$R_t^2 = 3.18$$

$$R_t^3 = 1.1 + 0.5(tan^{-1}(3*87.2))$$

$$R_t^3 = 1.88$$

$$R_t^4 = 3.1 + 0.5(tan^{-1}(3*86.7))$$

$$R_t^4 = 3.88$$

After $O_2$ is removed the reputations are again recalculated under the context $O_2$ was cached in which is also context 1. However, since the hit count of O2 is zero the reputations will stay the same without any increment.

$$R_t^1 = 2.0 + 0.5(tan^{-1}(0*60.1))$$

$$R_t^1 = 2.0$$

$$R_t^2 = 3.18 + 0.5(tan^{-1}(0*61.8))$$

$$R_t^2 = 3.18$$

$$R_t^3 = 1.88 + 0.5(tan^{-1}(0*0))$$

$$R_t^3 = 1.88$$

$$R_t^4 = 3.88 + 0.5(tan^{-1}(0*81.9))$$

$$R_t^4 = 3.88$$

Table 4-5: Updated reputation in all the known contexts for all registered sources

| No. | Name/Description | Reputation in context 1 | Reputation in context 2 |
|---|---|---|---|
| 1 | News service algorithm | 2.0 | 1.0 |
| 2 | Medical record Algorithm | 3.18 | 1.3 |
| 3 | Dependency graph based algorithm | 1.88 | 2.2 |
| 4 | PPM based algorithm | 3.88 | 1.2 |

**Adding new source/algorithm**

We now want to add a new algorithm specialized on the domain of an Enterprise Resource Planning (ERP) which the organization that owns the proxy server implemented to facilitate its daily rituals. The algorithms will predict future user access based on the semantic relationship pertinent to the ERP system.

When registering the new algorithm the Source Repository component will accept the new algorithm and its respective Context Profiler(s) (to be passed for the Context Manager) as parameters. Both of these (i.e., the algorithm and its Context Profiler) will be associated and registered by the components and will start participating in the prediction process immediately. For any algorithm to start this participation its reputation value will be set to one (the default value for a new algorithm).

# Chapter 5 - Implementation and Evaluation

## 5.1 Overview

In this Chapter we have presented a reference implementation of the proposed framework. We have used this implementation to evaluate the performance of the framework and subsequently we have also presented the result of this evaluation here.

Our primary goal with this implementation is to prove the possibility/feasibility of having such a framework and to set the baseline performance with a minimum possible configuration to serve as a means to gauge against for future improvements that would normally be a result of enhancements made to the various components of the framework.

Accordingly, we mainly implemented components that are newly proposed in this work while adapting a fairly trivial versions of components that either already exist as parts of other works (e.g., Cache Manager) or those components that are so complex by nature they need a separate study by themselves (e.g., Context Manager). For example, we did not set out to integrate the best possible Cache Manage in any way since it is a problem addressed by numerous others. As a result we settled for the default cache manager with LRU cache replacement policy.

In addition, we have made some assumptions while implementing and subsequently evaluating the proposed framework. We have presented these assumptions in the following section in detail with the reasons that lead to each of these inferences.

## 5.2 Assumptions

In this implementation we presuppose that different algorithms by themselves will have a significant difference in what they "see" from the same data source (i.e., user access log) to the point that having different algorithms that consume similar data source could serve as different Sources. Each source is expected to have a different view point of the system.

All components of the framework are assumed to be implemented at a single location but this assumption is of limited effect due to the basic nature of the framework that makes it location independent to function. The most significant effect of this assumption is having the cache accessible directly for the relevant components to write to. What is more, the cost in communicating with all sources is assumed to be similar hence all sources are contacted under all cases to provide their suggestions.

Since the framework does not put into consideration how suggestions are communicated, for our implementation we assumed the primary memory of the Aggregator is accessible directly by all sources. This way a Source will directly write its suggestion to the memory of the Aggregator. If the sources are assumed to be implemented in a remote location where the Aggregator's memory is not directly accessible we can add a component that will accept the suggestions from each Source and write it to the Aggregator's memory.

These assumptions will not affect the evaluations validity since its (i.e., the evaluation) primary objective is to quantify the predictive performance of this model compared to contemporary prefetching schemes which do not make use of collaboration.

## 5.3 Tools and Algorithms

In the evaluation, the model was implemented using Java 7 programming language with MySQL 5.5.34 backend to store generated rules and evaluation results. We choose Java programming language due its fair level of abstraction of unnecessary detail to facilitate rapid prototyping and testing of ideas as they arise which is greatly required in the experimental phase of this framework as compared to runtime efficiency which is of most demand when we actually integrate the framework into the production world.

We opted to store the generated rules and evaluation using a relational database (i.e., MySQL) due to its ability to support a fair level of querying out of the box (as compared to plain file systems) which are handy when we try to analyse the results of the evaluation.

The implementation incorporated three separate Sources (i.e., algorithms) namely $WM_O$ [93], Dependency Graph (DG) based [48] and Prediction by Partial Matching (PPM) [60] prefetching algorithms. $WM_O$ is a data mining based algorithm with modification to address issues pertinent to prefetching. The DG based scheme is a work similar to the one presented in [48]. The PPM algorithm we implemented here is similar to the one described in [60]. These algorithms mainly differ in how they count the frequency of document occurrence in the access log. A Detailed description of these algorithms is presented in Chapter 3 of this document, however the primary equations used by each of these algorithms is presented here as well.

**Dependency Graph**

For each transaction $T = \{p_1, p_2 ..... p_n\}$, Dependency Graph based approaches increase the frequencies of $p_i$ and $\{p_i, p_j\}$ for all $1 \leq i < j \leq n$ and $j - i \leq \min\{w, n\}$ where w is the look ahead window size. In this work we have evaluated the performance using multiple values for look ahead window sizes. No matter what value the look ahead window size is set to, however, DG will always update the frequency of only two entities in its database, one element with single item (i.e., $\{p_i\}$) and another that contains two items (i.e., $\{p_i, p_j\}$).

Hence, the database of a DG algorithm will have entries that consist of either a single web object or two regardless of the w value chosen. The pseudo code for the DG learning procedure that counts the frequency of items is presented in Figure 5-1, where D refers to the input database (i.e., list of transaction T) which serve as training data for the algorithm. W in this pseudo code refers to the look ahead window size.

$$P(p_j|p_i) = \frac{F(\{p_i, p_j\})}{F(\{p_i\})} \qquad (12)$$

Where $F(\{p_i, p_j\})$ and $F(\{p_i\})$ refer to the frequency count of $\{p_i, p_j\}$ and $p_i$, respectively.

DG then calculates the probability of a document $p_j$ being accessed given the document $p_i$ was accessed within the last w requests using Equation (12).

```
// Input: list of transaction
// Output: returns nothing but populates frequency count database
dgLearner(D):
BEGIN
      foreach T ∈ D:
            foreach j ← 0 to T.length:
                  increment frequency count of T[j] by one
                  for i ← j+1 to min(j + w, T.length):
                        increment frequency count of {T[j],T[i]} by 1

END
```

**Figure 5-1:** Pseudo code of DG learning algorithm

**Prediction-by-Partial Matching**

Prediction by Partial Matching will come with multiple orders, k. This value k is similar to w in DG but this time PPM will not only look ahead k requests as DG does with w but also updates the frequency of the various permutation of items in between as well. To put it more formally, for a transaction $T = \{p_1, p_2 \ldots p_n\}$ in the user access log a k-order PPM update the frequencies of $\{p_i\}$, $\{p_i, p_{i+1}\}$, $\{p_i, p_{i+2}\}$ ...., $\{p_i, p_{i+j}\}$ where $1 \leq i \leq n$, $i + j \leq n$ and $1 \leq j \leq k$. PPM then generates the probability of an object $p_{i+j}$ being accessed using Equation (13). The pseudo code for the PPM learning procedure that counts the frequency of items is presented in Figure 5-2, where k in this procedure is the order of the PPM model employed and D refers to the input database similar to the DG learning procedure in Figure 5-1.

$$P(p_{i+j}|p_i \ldots p_{i+j-1}) = \frac{F(\{p_i,\ldots p_{i+j-1},p_{i+j}\})}{F(\{p_i,\ldots,p_{i+j-1}\})} \quad (13)$$

Where F(x) refers to frequency count of x.

PPM differs from DG in the fact that how many frequencies are actually updated depends on the order k. Hence, the database of a PPM will have entries for elements containing one up to k items. What is more, even if we set k to one (i.e., 1-order PPM) the rules generated are different from a DG in that PPM will consider consecutiveness of the objects unlike DG. This is reflected in the pseudo code presented in Figure 5-2 on the line that increments the frequency.

```
// Input: list of transaction
// Output: returns nothing but populates frequency count database
ppmLearner(D):
BEGIN
 foreach T ∈ D:
  foreach j ← 0 to T.length:
   increment frequency count of T[j] by one
     for i ← j+1 to min(j + k, T.length):
        increment frequency of{T[j],T[j+1],…T[i-1],T[i]} by 1
END
```

**Figure 5-2:** Pseudo code of PPM learning algorithm

**WM$_O$**

As stated previously, WM$_O$ is a modification of contemporary data mining algorithms, in particular Apriori, to address the issue of order that should be maintained in prefetching area. Since Apriori was initially designed to address bucket mining problem in a shopping scenario where it only considers which objects are purchased together without consideration of order of purchase. Hence WM$_O$ is a modification to the candidate generation process of the Apriori algorithm as presented in Figure 5-3. In this procedure, L$_k$ is the set of frequent items of size k-1 where k is the size of candidates we want to generate and F$_1$ is the set of all frequent items of size one. Here when we say S ≺ C, it means S is a subsequence of C.

```
// Input:  set of frequent items of size k-1
//         set of all frequent items of size 1
// Output: candidate item sets of size k
generateCandidate(L_k, F_1):
BEGIN
   Candidates ← ∅
   foreach l ∈ L_k:
      foreach v ∈ F_1:
         if v ∉ l:
            C ← {l_1,…, l_k, v}
            if ∀ S ≺ C ⇒ S ∈ L_k:
               insert C to candidates
   return Candidates
END
```

**Figure 5-3:** Pseudo code of candidate generation algorithm

All of the above stated algorithms use the same set of data sources, however, are expected to come up with their own independent set of prefetching rules. This is based on our assumption in the previous section that states "different algorithms will 'see' different patterns even if the data sources they use is similar". Of course the information they see will be even more diverse if they are exposed to data at the different segments/locations of the request-response chain due to the full view they are able to get if such a scheme is deployed. However, by making all use identical data source, it can be seen how much difference each algorithm by itself can make in the overall multi-source scheme.

In other words, basing all algorithms on a single data source shows the actual improvement made by using multiple sources without mixing it to the improvement seen due to having multiple data sources (e.g., access logs from the different levels of the request chain). Having multiple data sources will obviously improve the overall predication performance of a scheme, given that these data sources possess relevant information, even if all these data sources are explored by a single algorithm. We can safely conclude this fact is true based on the very basic definition of all learning algorithms (which prediction algorithms are part of) that says a learning algorithm will improve its performance as it gets more and more learning data.

## 5.4 Framework Components Implementation

**Request Agent**

As discussed previously the Request Agent component is responsible for accepting incoming requests from clients and aims to hide how requests arrive and where they come from (i.e., local client running on the same machine as the framework or a remote standalone client that sends its request over the network). In this implementation, we focused on realizing a part of the component that accepts requests from a local client. The reason for doing so is the fact that issues for the client handler to address are more or less similar regardless of the source of request with the exception of some additional requirement when we consider communication of requests over network, which lays beyond the scope of the framework.

```
// Input:  Id of the object requested by the client
//         Id of the client to return the object to
// Output: the requested object
acceptRequest(o,client):
BEGIN
   If O in cache:
      object ← cache.getObject(o)
   else:
       // Pending request is the list of all requests by all clients
      PendingRequests.add(o,client)
      FatchAgent.fetch(o)
      Aggregator.informContextChange()
   informContextManger()
END
```

**Figure 5-4:** Pseudo code of accept request procedure

The accept request functionality of Request Agent we implemented is presented in Figure 5-4. In the procedure presented the parameter $O_{id}$ refers to the id the object requested. The accept request procedure is invoked whenever a new request arrives from clients and the request information is passed to the procedure as a parameter. Following our assumption that the cache is accessible by the framework, when Fetch Agent's fetch method is invoked it will store the object to cache and call the notify procedure of the Request Agent with the object id as a parameter. It can be observed from Figure 5-4 that the request will inform context change only when there is a cache miss, which in turn may result in recalculation of prefetch queue. However, in production implementations a separate process may initiate this recalculation periodically even without a cache miss.

**Cache Manager**

To address the objectives of this evaluation we have implemented a minimal cache manager with a least recently used cache replacement policy. The policy removes objects which were not accessed for a while whenever there is a cache space shortage. To avoid removing objects which just got cached even before being accessed for the first time, we have included time of caching to this calculation.

```
// Input:  Object to store to cache
// Output: returns nothing
storingObject(element):
BEGIN
   if #elementsInCache > cacheLimit:
       LRU()
   cache.cachedOn ← now
   cache.put(element)
   #elementsInCache += 1
END
```

**Figure 5-5:** Pseudo code of storing object to cache

```
// Input:  Noting
// Output: returns nothing but removes an object from cache
LRU():
BEGIN
   v ← Δ(cache[0].cachedOn)/ Δ(cache[0].lastHit)
   i ← 0
   foreach j ← 1 to Cache.length:
      c ← Δ(cache[j].cachedOn)/ Δ(cache[j].lastHit)
      if x > v:
            i ← j
            v ← c
   if cache[i].prefetched == true:
         updateReputations(cache[i])
```

**Figure 5-6:** Pseudo code of removing objects from cache

The pseudo code for the cache removal procedure (i.e., LRU procedure) is presented in Figure 5-6, which is invoked by the store to cache procedure presented in Figure 5-5. In Figure 5-6 Δ(cache[x].cachedOn) and Δ(cache[x].lastHit) refer to the difference in time from now and the time object x was cached and last accessed, respectively.

As it can be observed from the presented procedures we have implemented a Cache Manager which removes objects exactly when there is a shortage of cache space (i.e., the cache removal procedure is invoked when shortage of cache space is observed while storing objects to cache). What this means is least recently used objects are removed right at the time when new objects are to be saved. However, in real world scenarios having a separate process that periodically checks if the cache is full (or nearly full) may avoid the delay due to full caches when we try to save new objects by performing this task in parallel to other happenings such as fetching objects rather than as part of the sequential steps of storing objects to cache.

**Context Manager**

In our implementation we were not able to employee a complex Context Manager because it would shift our focus from the primary objective of this evaluation since context management is by itself a sophisticated issue that needs separate addressing. Accordingly, the context management we used is a session level processing. In the training and evaluation data we have marked each session with a numerical value that represents each of the contexts we presumed to have. Hence the context profiler reads this values to find out the context for a given session. We by no means claim this to be a replacement for a full-fledged Context Manager but we think it shows an abstract view of the basic workflow involved in the various processes of the Context Manager for further studies to base of.

**Reputation Manager**

The Reputation Manager, in particular the updateReputation functionality, is presented in Figure 5-7. This procedure is initiated when an object is removed from cache hence the object removed is passed as a parameter to it. Notice, however, this procedure is invoked by Cache Manager only when prefetched objects are removed from cache.

```
// Input:   the object to be remove from cache
// Output: returns nothing but update reputation of sources
updateReputations(o):
BEGIN
  For j ← 0 to SR.length:
    SR[j][o.context] += SR[j].getReputation[o.context] +
                δ *tan⁻¹(o.hitCount * o.confidences[j])

END
```

**Figure 5-7:** Pseudo code of updating reputation of sources

**Fetch Agent**

```
// Input:  Id of the object to fetch
//         the initiator of this fetch (i.e., ordinary, prefetch)
// Output: stores the object to cache and notifies the requester
fetch(O_id, caller ← "ordinary"):
BEGIN
   Object ← transaction [O_id]
   if caller == "prefetch" :
      object.prefetched ← true
      Cache.storingObject(object)
      PrefetchAgent.notify(O_id, Object)
   else:
      object.prefetched ← false
      Cache.storingObject(object)
      RequestAgent.notify(O_id, Object)
END
```

**Figure 5-8:** Pseudo code of fetch procedure of the Fetch Agent

Since this evaluation is based on a log file input the Fetch Agent we have implemented will return whatever the next entry is in the evaluation transaction which is the object requested by the client. This process is presented by the fetch procedure of the pseudo code presented in Figure 5-8. Here the object id is passed as a parameter and the procedure returns the actual web object. In this case since we assumed the client's cache is accessible the fetch procedure simply writes it to the client's cache. What is more, caller is an optional value that will tell the fetch procedure the initiator of this fetch, by default it is assumed to be an ordinary request but it could be a prefetch.

```
// Input:  Nothing
// Output: Returns nothing but fetch objects to cache that
//         are listed in the prefetch queue
prefetch():
BEGIN
   For j ← 0 to prefetchQueue.length:
      // Pending prefetches is the list of all prefetch requests
      PendingPrefetches.add(prefetchQueue[j])
      FatchAgent.fetch(o, "prefetch")
END
```

**Figure 5-9:** Procedure prefetching an object

**Prefetching Agent**

The Prefetching Agent in this implementation will go through the prefetch queue sequentially and calls the Fetch Agent without additional hassle due to the assumption we made in a previous section that basically omits the need to consult the Prefetch Watchdog. Once the Fetch Agent brings the requested object to cache, it will inform the prefetch agent about it with the Prefetch Agent's notify procedure to add prefetching related information to the object. The algorithm of the Prefetching Agent that actually requests this fetch is presented in Figure 5-9.

**Prefetch Watchdog**

For this evaluation we did not implement a Prefetch Watchdog to minimize complexity. What this means is perfecting request can be entertained at any time without consideration to resources such as bandwidth. However, in production implantations this is not feasible (i.e., sidestepping consideration of resource availability while performing prefetching), hence a mechanism to determine the best time to prefetch is necessary. Accordingly, this issue is addressed in many other schemes, thus could easily be adopted from such works.

**Source Repository**

The source repository employed in our implementation is a simple list of Source objects which consisted of functionalities to add and remove sources dynamically. Since we have not implemented a complex context manager the process of adding and removing sources to the repository is not that much different than simply adding an object to the data structure underneath the repository. This is because no additional step is undertaken to register context profilers when we add a new source and unregister it when we remove the respective source.

**Aggregator**

The Aggregator component we have implemented accepts suggestions from the all Sources we have registered and creates the aggregate list to filter objects using the reputation of each source. One important thing to notice about our implementation is that we have not considered the cost associated with each source to be different. This is based on our assumption presented earlier to minimize complexity of implementation at this point. The procedure for performing this is presented in Figure 5-10. Here SR is the set of all registered Sources and current context to be an object that represents the current context information that is deemed relevant for predicting upcoming user access.

```
// Input:  Nothing
// Output: Returns noting but updates the prefetch queue
updatePrefetchQueue():
BEGIN
  aggregateList ← ∅
  currentContext ← ContextManager.getCurrentContext()
  for j ← 0 to SR.length:
     c ← acceptSuggestions(SR[j],p, currentContext)
     foreach e ∈ c:
       if e ∉ aggregateList:
          insert e in to aggregateList
       else:
          aggregateList[e].setConfidence(j,e.confidence[j])

  // calculate total confidence for each of the suggested objects
  foreach o ∈ aggregateList:
     o.totalConfidence = 0
     for k ← 0 to SR.length:
      o.totalConfidence += o.confidence[j] * SR[k].getRepuation()

  // sort based on totalconfidece in a descending order
  sort(aggregateList)
  return aggregateList.sublist(0, p)

END
```

**Figure 5-10:** Procedure for querying suggestion from each source

## 5.5 Evaluation

**Data Generation**

For the evaluation purpose we have generated synthetic data which is merely a list of web transactions. The primary reason for using synthetic data is the fact that we need the input data to comprise of multiple contexts explicitly specified due to the simplistic view of our implementation of the Context Manager as stated in the previous section. This simply is not available in real production logs. Hence, the data generation process accepts the presumed number of documents on the server as parameter and will create two sets of transactions each representing a unique context.

The first context is a generic web content whereby navigation is based on hyperlinks. That means a user viewing a document could only navigate to documents that are referenced by this document via a hyperlink connection. Navigations that do not follow this connection are often due to bookmarked links the user follows by directly typing in a URL which in both cases is practically impossible to predict the flow.

To generate transactions in this context, first documents in the site are connected to each other randomly to represent a connection on hypertext documents. The connections formed are directed, which in other words means having document A connected to B doesn't necessarily imply B is connected to A. Hence, users can navigate from A to B but the reverse may not be possible. Accordingly when the transactions are generated this directed connection is considered.

Once the random connection process completes, if a document is found to be not connected to any other document at all it will be connected to the document with the maximum number of outgoing connections. This is to mimic the fact that some documents, especially homepages of a website, are more or less connected to almost all documents within the site.

When producing a transaction we started with a random document and then assume the user navigates randomly to one of the documents connected to the current document. This is in light of the fact that the document a user first requests from a site is purely random from the provider's point view [94]. However, once a user retrieves and views a web document s/he will navigate to one of the documents that are listed to be connections for the document s/he is currently viewing. Among these set of documents (i.e., documents that are referenced by the document currently being viewed), the user could choose to navigate to any randomly. This form of data generation is similar to the one presented in [93]. We have presented the pseudo code for this data generation procedure in Figure 5-11.

The second context is where users navigate through a site without a predefined document connection. That is, the navigation is assumed to be guided by a conceptual connection that looks random from the hyperlink point of view. What this means is at any single point a user could navigate to any other document on the site without the need to follow any predefined reference relationship. This is possible if the application provides some sort of semantic relationship which guides the user navigation. We have presented the pseudo code for this data generation procedure in Figure 5-12.

Transactions are generated with the length of each transaction being random between a minimum, minTrans, and maximum, maxTrans. Both the actual number of transactions and the minimum and maximum length of a transaction are parameters set at the beginning of the data generation process according to Table 5.1.

```
// Input:   Total number of document in the transaction
//          Number of transactions to generate
//             Minimum size of a transaction
//             Maximum size of a transaction
// Output: List of transactions
generate(noOfDocs, numOfTransactions, minTransSize, maxTransSize):
BEGIN
   // Generate Mappings
   Output ← ∅
   documents ← objects[noOfDocs]

   for i ← 0 to documents.length:
      for j ← 0 to random(1, noOfDocs - 1):
         RANDOM: mappingDoc ← documents [random()]
         if mapingDoc == documents[i]:
            do not increment j
            goto RANDOM
         else:
            documents[i].mapTo(mappingDoc)
   // If the document is not connected to anything
   // connect to the most popular document
   for i ← 0 to documents.length:
      if documents[i].countOfIncomingReferences == 0
         dcouments[popular]. mapTo(documents[i])
   // Generate Transactions
   for k ← 0 to numOfTransactions:
      noOfTrans ← random(minTransSize, maxTransSize)
      currentDoc = documents[random()]
      newTransaction ← ∅
      for l ← 0 to noOfTrans:
          REDO: nextDoc ← currentDoc.mapings[random()]
          if nextDoc ∈ transaction:
             do not increment l
             goto REDO
          else:
             transaction.add(nextDoc)
             currentDoc ← nextDoc
      output.add(newTransaction)

END
```

**Figure 5-11:** Procedure for generating synthetic data in the first context

```
// Input:   Total number of document in the transaction
//          Number of transactions to generate
//          Minimum size of a transaction
//          Maximum size of a transaction
// Output: List of transactions
generate(noOfDocs, numOfTransactions, minTransSize, maxTransSize):
BEGIN
   Output ← ∅
   documents ← objects[noOfDocs]

   // No Mapping needed
   // Generate Transactions
   for k ← 0 to numOfTransactions:
      noOfTrans ← random(minTransSize, maxTransSize)
      currentDoc = documents[random()]
      newTransaction ← ∅
      for l ← 0 to noOfTrans:
         REDO: nextDoc ← documents[random()]
         if nextDoc ∈ transaction:
            do not increment l
            goto REDO
         else:
            transaction.add(nextDoc)
            currentDoc ← nextDoc
      output.add(newTransaction)

END
```

**Figure 5-12:** Procedure for generating synthetic data in the second context

**Table 5-1:** Syntactic Data generation parameter and values

| Parameter | Value |
| --- | --- |
| Number of documents | 20 |
| Minimum length of a transaction | 4 |
| Maximum length of a transaction | 15 |
| Number of transactions | 10,000 |

**Configurations**

In addition to the various parameters required for the data generator, the framework's implementation consists of a number of configuration values. This parameters are the look ahead window size w in DG, the order value for PPM k, support and confidence values for $WM_O$ algorithm, cache size, learning rate δ, and prefetch length p. We have experimented with various values for each in this evaluation as it can be seen in subsequent section and Table 5-2 which presents the range of values used for each parameter.

**Table 5-2:** Framework parameters and the range of values used for evaluation

| Parameter | Values |
|---|---|
| Look ahead window size W | 1, 2, 3, 4 |
| PPM order K | 1, 2, 3, 4 |
| Support | 0.5, 1, 1.5, 2, 2.5 % |
| Confidence | 0.1, 0.3, 0.5, 0.8 |
| Cache size | 10, 30, 50, 80, 100 |
| Learning Rate δ | 0.1, 0.3, 0.5, 0.8, 1 |
| Prefetch Length p | 1, 3, 7, 10 |

**Measurement Criteria**

In the evaluation we conducted we used efficiency or Hit per Waste Ratio (HW) as defined in Chapter 2 of this document. We chose efficiency as the primary criterion to gauge performance of this framework because it encompasses other measures defined (i.e., Hit Ratio and Waste Ratio) and it provides an insight into the framework's applicability in domains where devices have limited resources which is basically all contemporary mobile client end devices. We chose Hit Ratio over Byte Hit Ratio and Waste Ratio over Byte Waste Ratio while calculating efficiency since the input data we used does not possess size information with the aim of keeping the complexity to a minimum as our primary objective is to measure prediction performance.

## 5.6 Findings

After a successful evaluation of the scheme we were able to find out that the scheme proposed is much more efficient than the three algorithms chosen for comparison employed individual. These algorithms are not only used for weighing the scheme against, but also make up the scheme constructed for the evaluation as well. In other works, the scheme defined is gauged against its individual building blocks which in turn represents the existing algorithms in the prefetching world. As presented in Figures 5-13 to 5-31, the proposed scheme is much more efficient (i.e., imposes minimal wastage or overhead) with a fair level of effectiveness.

**Comparison against Dependency Graph based approaches**

In Figures 5-13 to 5-15 we have presented the comparison of the efficiency of the proposed framework with a dependency graph based approach presented earlier. We measured efficiency using the definition presented in Chapter 2 in Equation (6). The figures show this contrast at varying levels of look ahead window sizes. Figure 5-13 focuses on the minimum performance observed for other parameters presented in Table 5-2, while Figures 5-14 and 5-15 highlight the average and maximum performances observed, respectively. In Figures 5-16 and 5-17 we have presented the comparison with varying prefetch lengths and cache sizes.

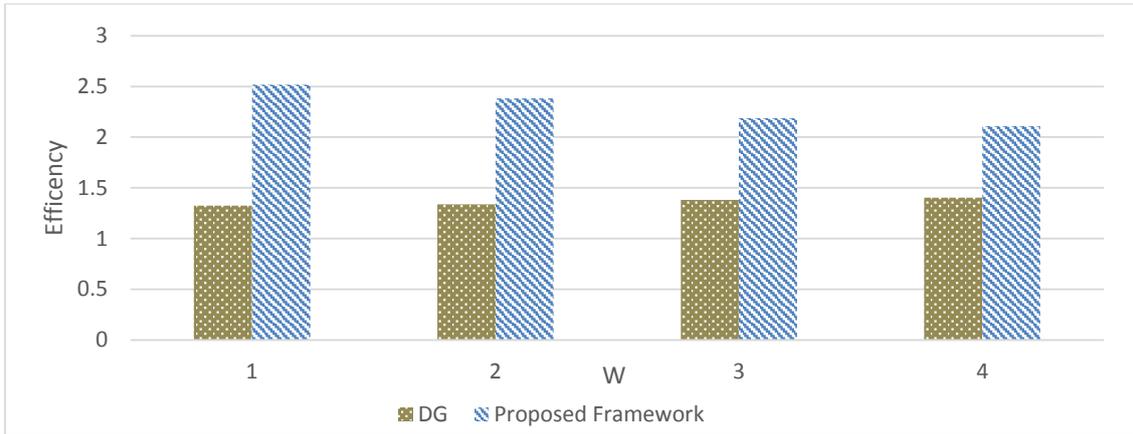

**Figure 5-13:** Comparison of minimum efficiency of the proposed framework vs. Dependency Graph based scheme over varying look ahead window sizes (w)

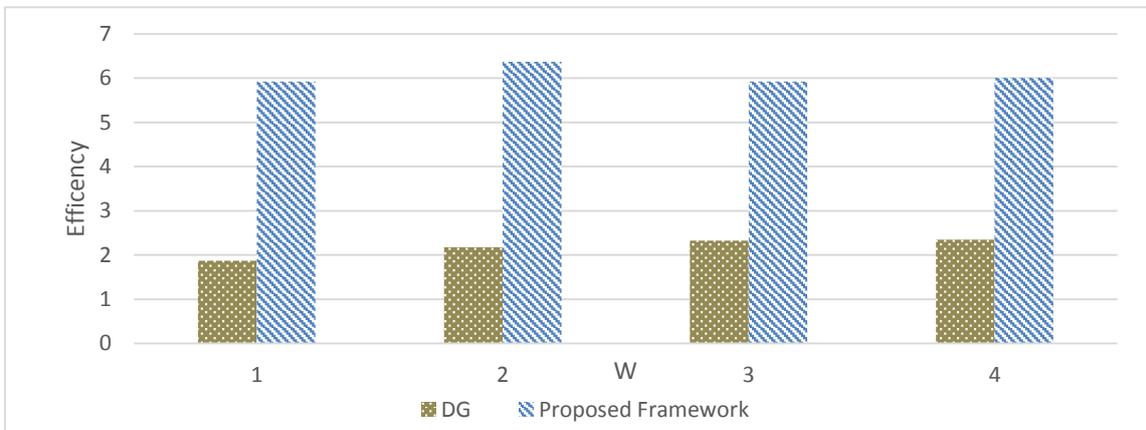

**Figure 5-14:** Comparison of average efficiency of the proposed framework vs. Dependency Graph based scheme over varying look ahead window sizes (w)

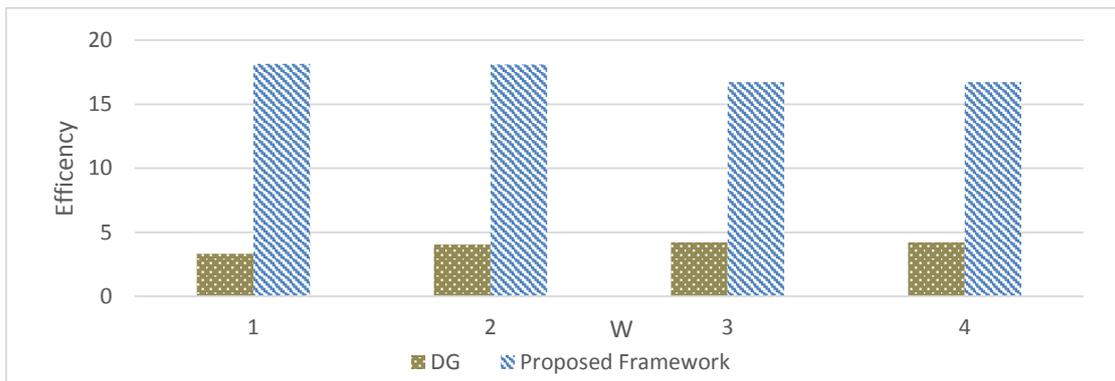

**Figure 5-15:** Comparison of maximum efficiency of the proposed framework vs. Dependency Graph based scheme over varying look ahead window sizes (w)

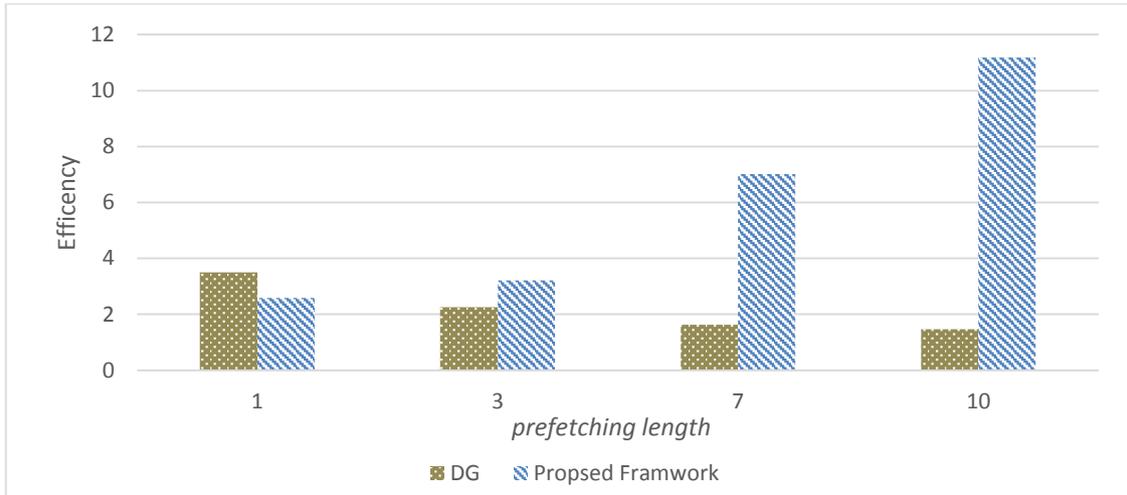

**Figure 5-16:** Comparison of average efficiency of the proposed framework vs. Dependency Graph based scheme over varying prefetching length

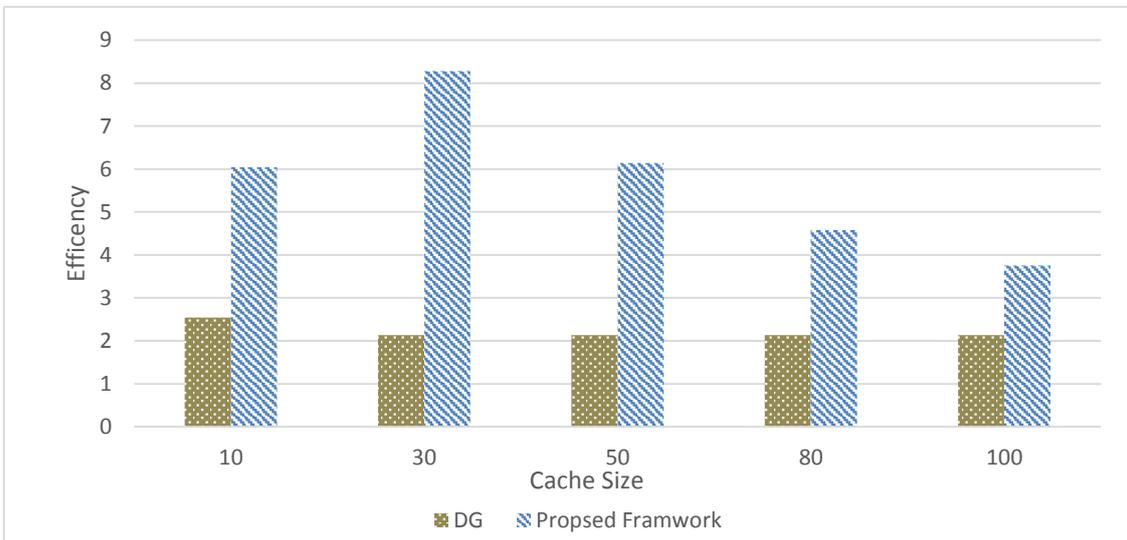

**Figure 5-17:** Comparison of average efficiency of the proposed framework vs. Dependency Graph based scheme over varying cache sizes

**Comparison with PPM based approach**

In Figures 5-18 to 5-20 comparison of the efficiency of the proposed framework with a prediction by partial matching scheme is presented. The figures show this contrast with different order PPM schemes again with minimum, average and maximum performance for parameters other than order. In all of these figures, it can be observed that efficiency of first order PPM is not defined since in our evaluation PPM was unable to predict any item resulting in zero number of prefetch which in turn resulted in zero wastage. In Figures 5-21 and 5-22, similar to the Dependency Graph comparison, we have presented the assessment with varying prefetch lengths and cache size.

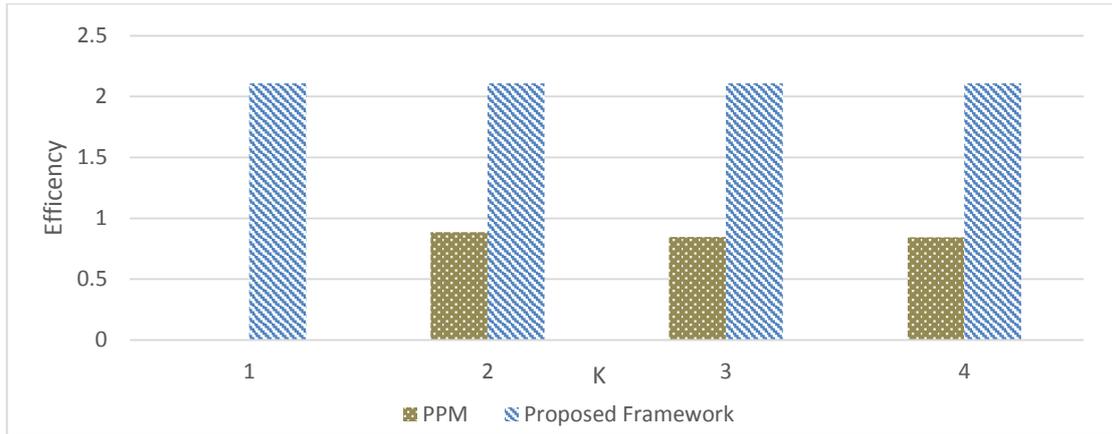

**Figure 5-18:** Comparison of minimum efficiency of the proposed framework vs. Prediction by Partial Matching scheme over varying orders (k)

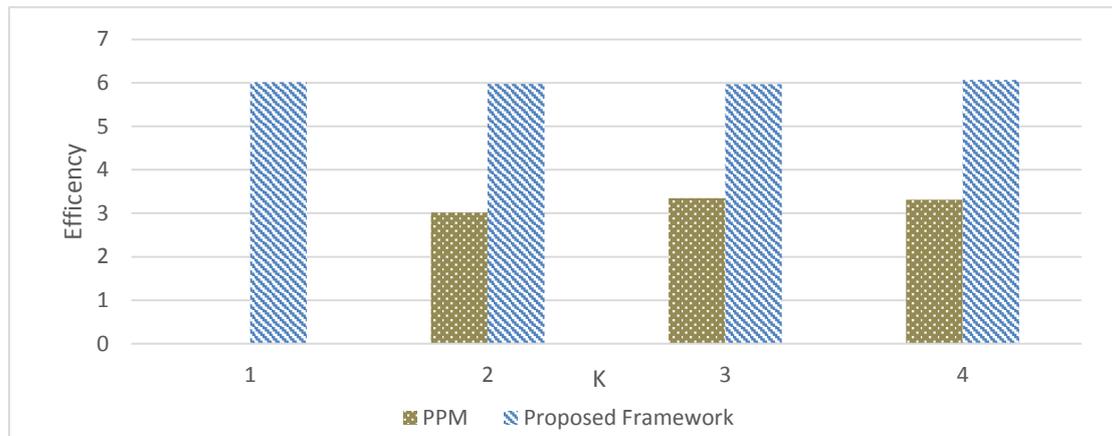

**Figure 5-19:** Comparison of average efficiency of the proposed framework vs. Prediction by Partial Matching scheme over varying orders (k)

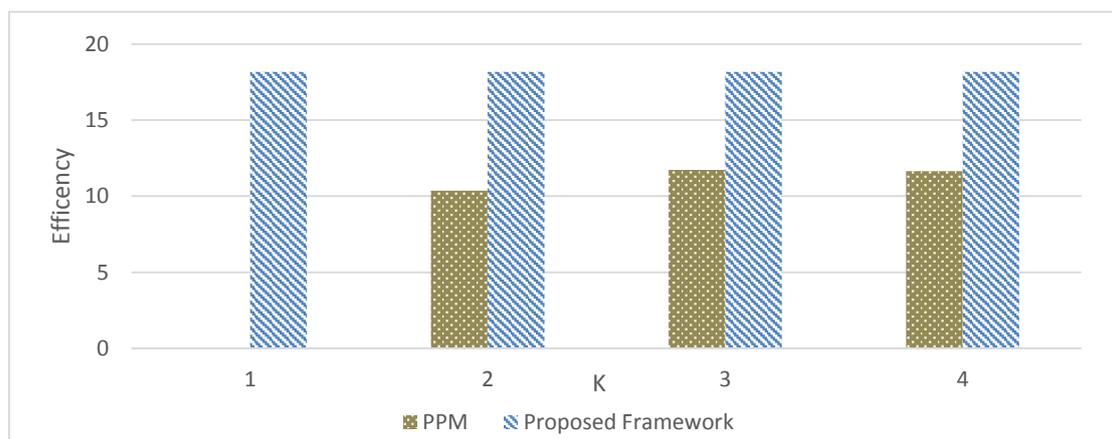

**Figure 5-20:** Comparison of maximum efficiency of the proposed framework vs. Prediction by Partial Matching scheme over varying orders (k)

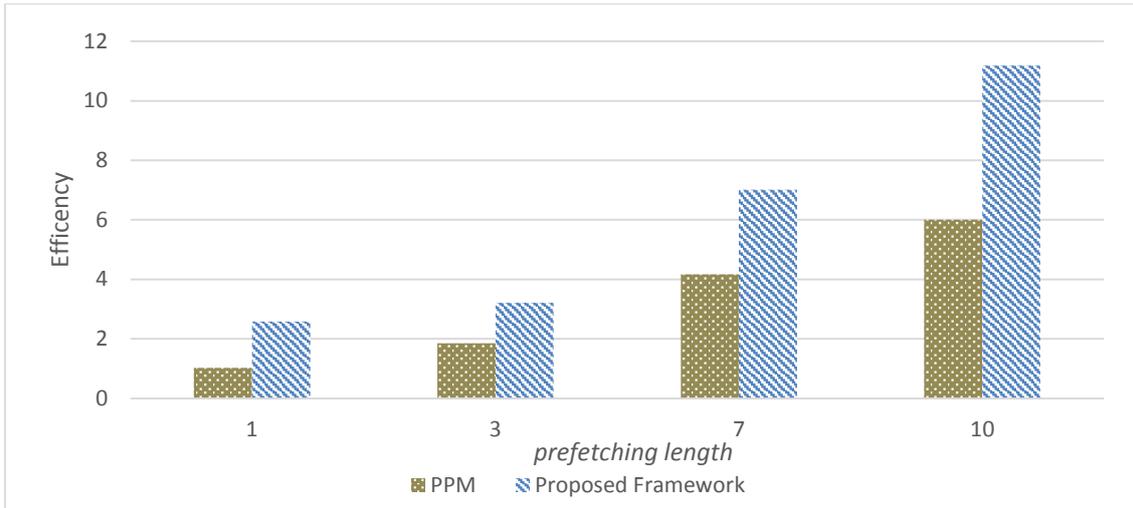

**Figure 5-21:** Comparison of average efficiency of the proposed framework vs. Prediction by Partial Matching scheme over varying prefetching length

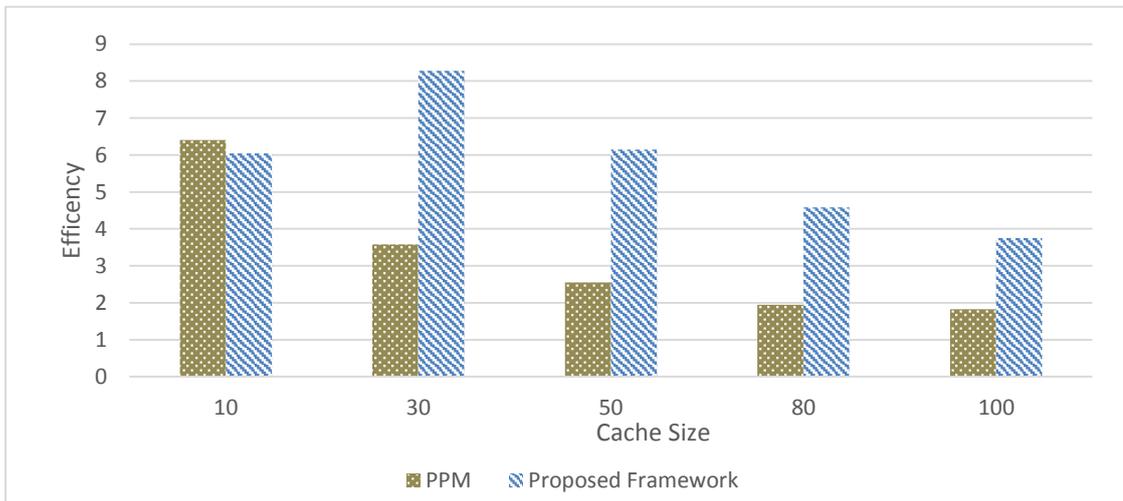

**Figure 5-22:** Comparison of average efficiency of the proposed framework vs. Prediction by Partial Matching scheme over varying cache sizes

**Comparison with WM$_O$ Algorithm**

Comparison of the efficiency of the proposed framework with the WM$_O$ scheme is presented in Figures 5-23 to 5-28. Since WM$_O$ is a data mining based scheme that mainly relies on two parameters that were not available in DG and PPM schemes, namely Support and Confidence. The first three sets of figures show the comparison with different support values while the remaining three show comparisons with different confidence values. In Figures 5-29 and 5-30 we have presented the assessment with varying prefetch lengths and cache size.

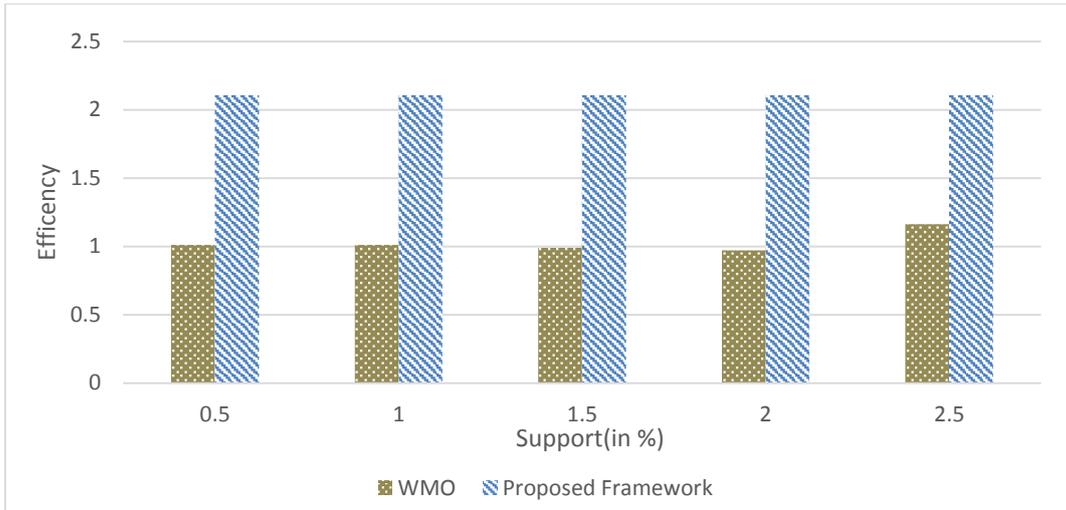

**Figure 5-23:** Comparison of minimum efficiency of the proposed framework vs. $WM_O$ scheme over varying support values

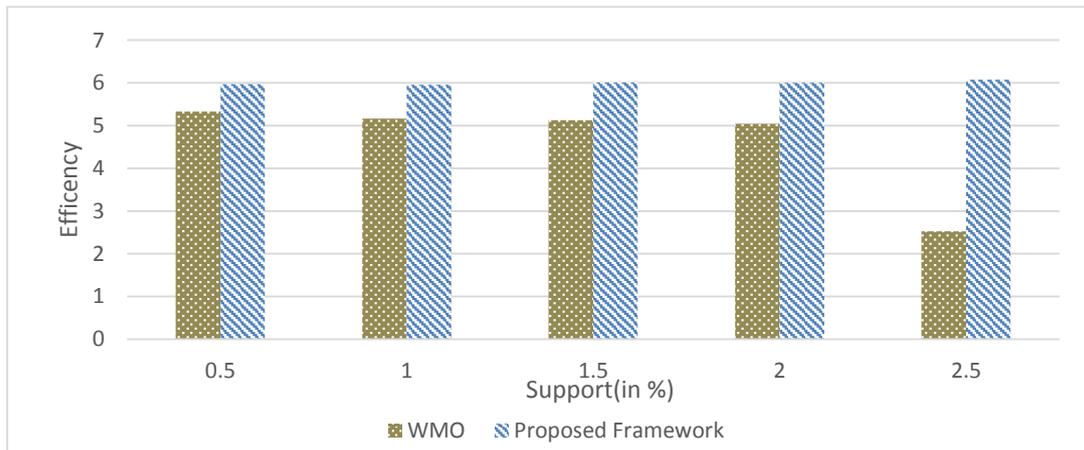

**Figure 5-24:** Comparison of average efficiency of the proposed framework vs. $WM_O$ scheme over varying support values

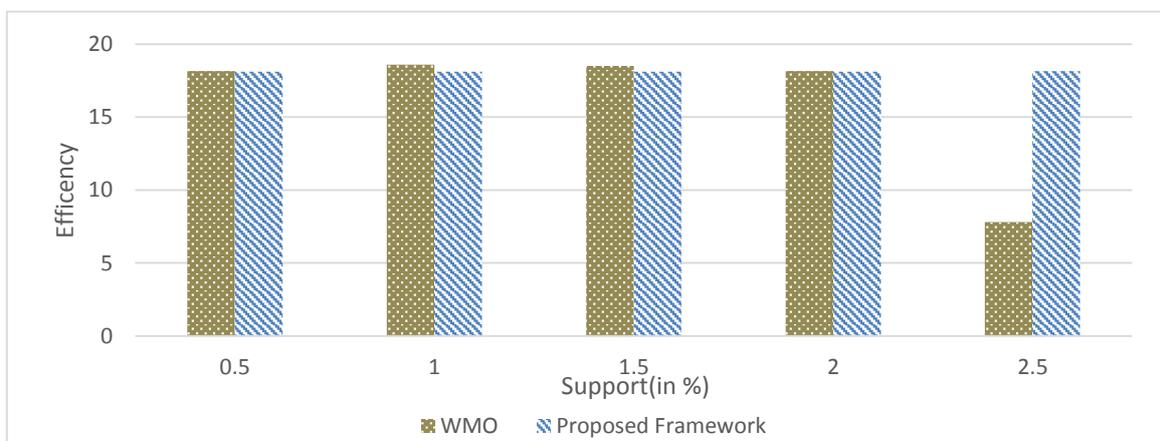

**Figure 5-25:** Comparison of maximum efficiency of the proposed framework vs. $WM_O$ scheme over varying support values

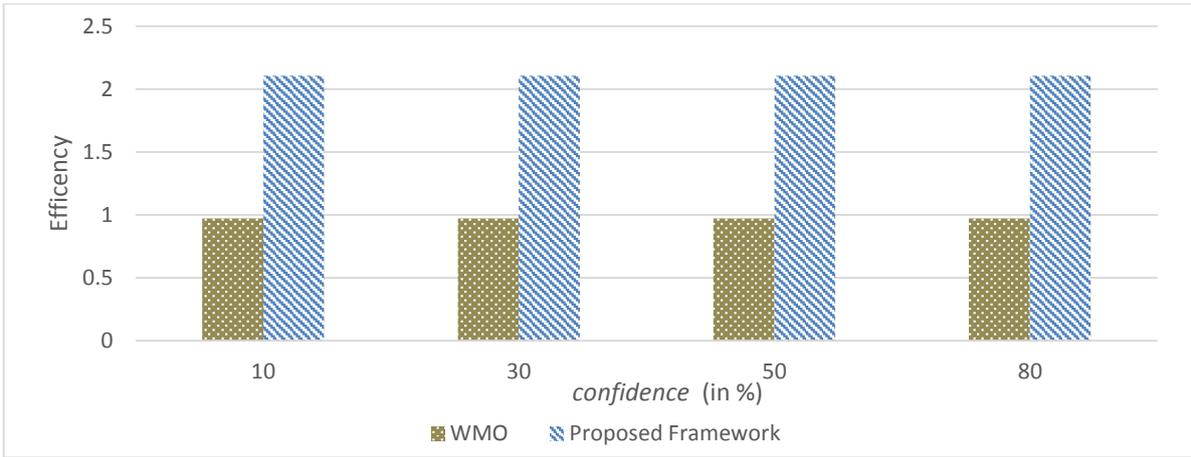

**Figure 5-26:** Comparison of minimum efficiency of the proposed framework vs. $WM_O$ scheme over varying confidence values

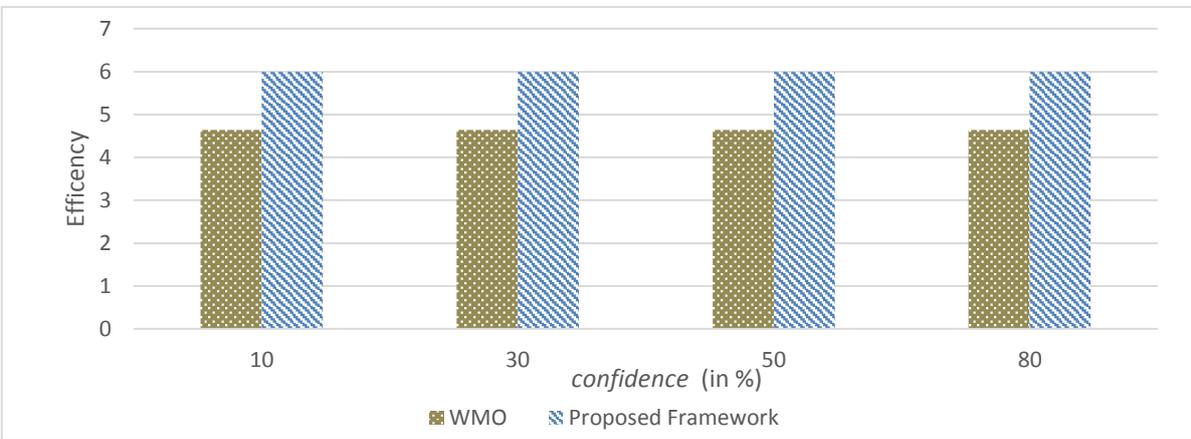

**Figure 5-27:** Comparison of average efficiency of the proposed framework vs. $WM_O$ scheme over varying confidence values

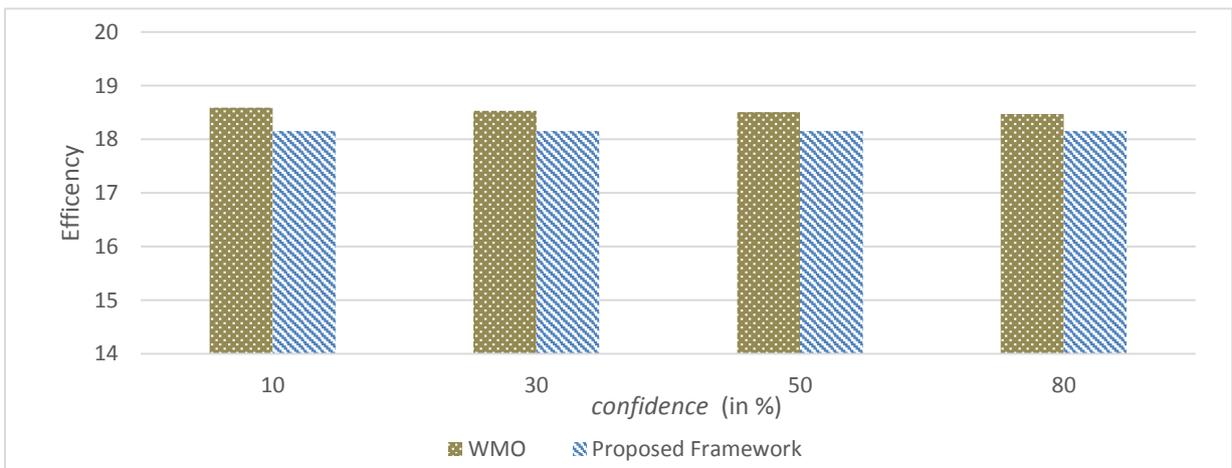

**Figure 5-28:** Comparison of maximum efficiency of the proposed framework vs. $WM_O$ scheme over varying confidence values

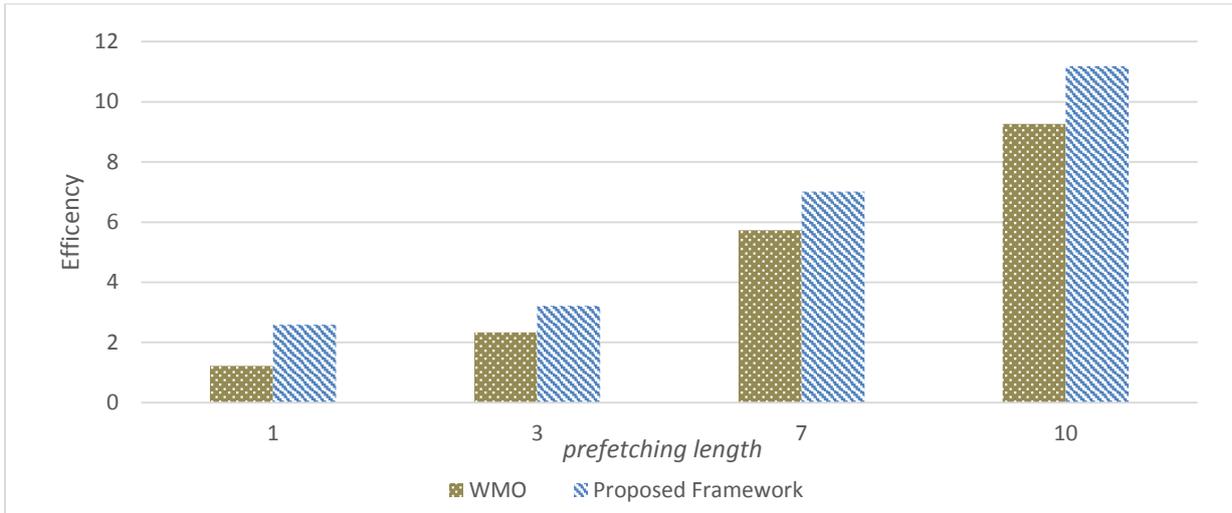

**Figure 5-29:** Comparison of average efficiency of the proposed framework vs. WM$_O$ scheme *over varying prefetching length (δ)*

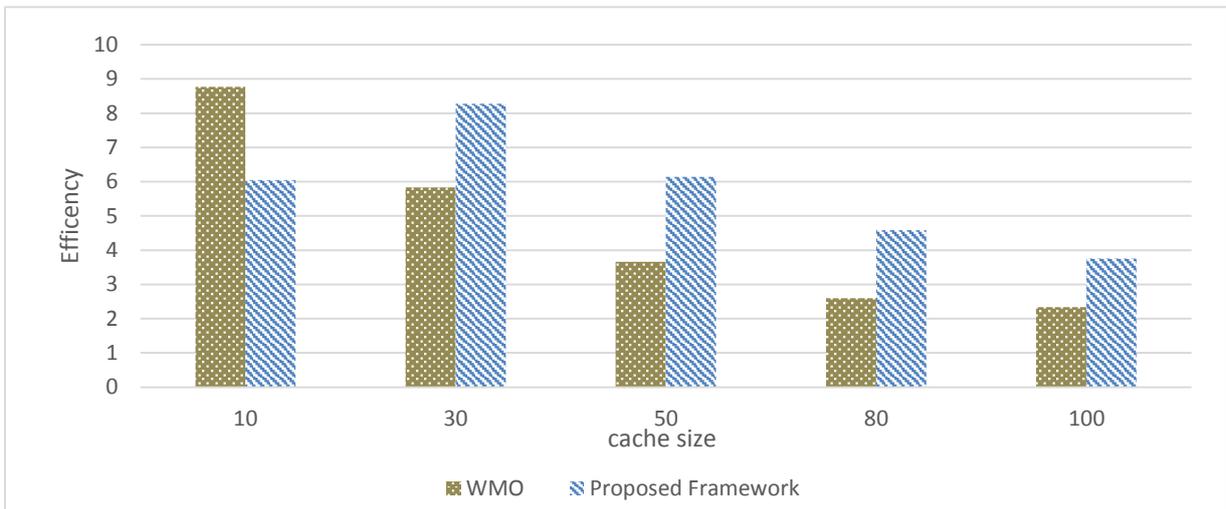

**Figure 5-30:** Comparison of average efficiency of the proposed framework vs. WM$_O$ scheme over varying cache sizes

**Effect of Learning Rate**

Finally we have examined the performance of the proposed framework over a range of learning rate values as presented in Figure 5-31. This figure depicts the minimum, maximum and average performances of the framework at a number of learning rate values.

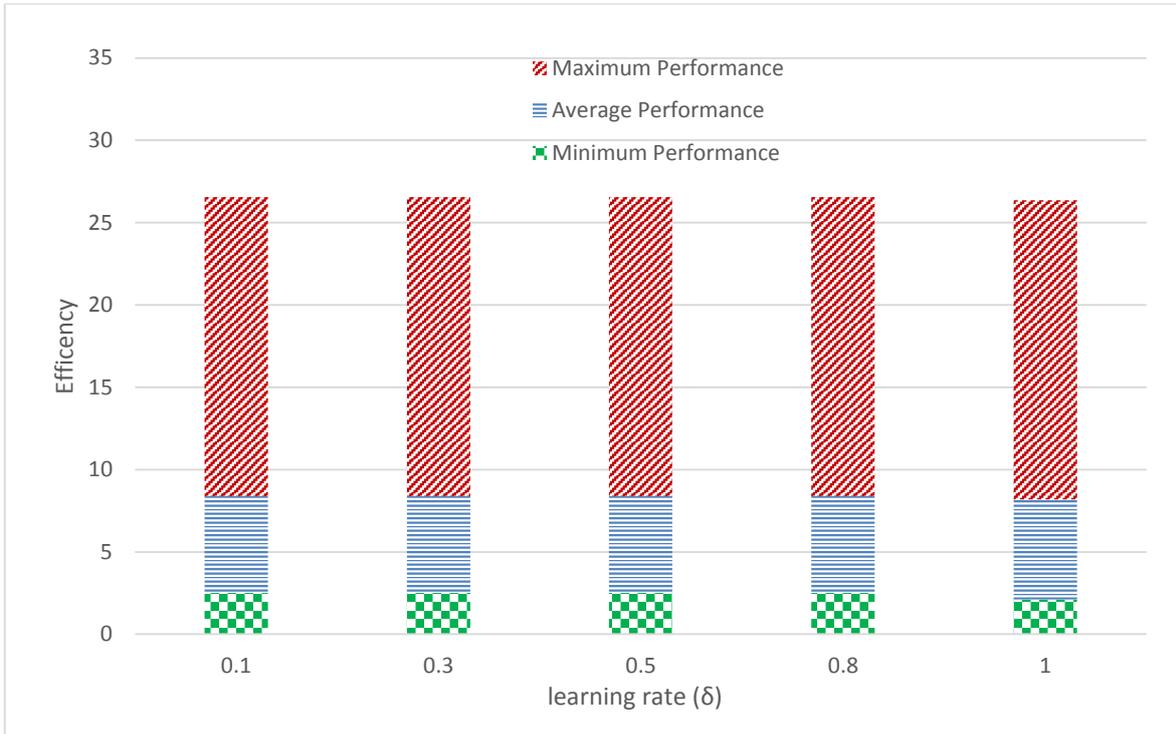

**Figure 5-31:** *Efficiency of the proposed framework over varying learning rate (δ)*

# Chapter 6 - Discussion

Based on the findings observed in the previous Chapter we present here our perception of the result in relation to the research objectives stated in Chapter 1. As a reminder, the primary objective of this work was to design a framework that enables transparent multi-source prefetching and answers the questions

- How best will the use of multi-source prefetching with adaptive weights improve the predictive performance of prefetching schemes?
- How best will the use of multi-source prefetching with adaptive weights address the issues that arise from changes in the cyber world trends, in particular changes in user end device types?

In addition, we have defined efficiency to be the total work done (i.e., successful predictions) per the total cost (i.e., unsuccessful predictions). Schemes that perform well in this measure are highly suitable for resource constrained devices due to the focus on minimizing wastage.

Accordingly, the findings we have observed are encouraging considering the limited setting we have employed. To begin with, we have not implemented complex algorithms for many of the components within the framework (e.g., context manager, cache manager). Choosing the right algorithm for each of these components requires a separate research. However, even with the simple algorithms employed the prediction performance observed outperformed all of the schemes adopted for comparison.

The fact that the scheme performed well in terms of efficiency compared to the three contemporary algorithms employed separately implies it is less aggressive than its counterparts. Baring in mind that aggressiveness is unacceptable especially in resource constrained environments (e.g., mobile user end devices we are accustomed to these days), it can be projected that the newly proposed scheme is more suitable for such devices than the others mentioned.

What is more, the data we have used for all algorithms is identical but the combination of algorithms by itself had an impact on the final prediction performance. Moreover, the designers of $WM_O$ algorithm claim it to be a generalization of the other two (i.e., PPM and DG) which theoretically means the prediction it produces should highly resemble the latter two. What this implies is that the prediction performance should have not changed by much when combining the three algorithms (i.e., $WM_O$ 'see' what the other two are able to see) since, supposedly, they all predict the same objects. However, the finding shows that the combination has a major effect even if the data source is the same and the algorithms are similar.

When we compare the proposed schemes' performance over varying prefetching lengths, it can be observed that DG and PPM tend to outperform the proposed scheme with a single object to prefetch but as we prefetch multiple items the performance of contemporary algorithms tends to diminish while that of the proposed framework's performance grows. Having a scheme that predicts multiple object ahead of time well is desirable since prefetching multiple items successfully at a time is obviously much more rewarding than prefetching a single item as it can be seen in all figures that show performance over varying prefetch length (i.e., Figure 5-19, Figure 5-21, and Figure 5-29). When we see the comparison of the framework with $WM_O$ over varying prefetch length the $WM_O$'s performance grows with increasing prefetching length, however, the framework still performs better than $WM_O$.

When we consider cache size, we have found cache size of 30 elements to be the ideal cache size with relatively optimal performance among those we tested with. However, we have not experimented much with the Cache Manager in our framework to say much about this in a conclusive manner. Yet, one obvious explanation for the poor performance with small cache sizes is that the scheme was unable to maintain fetched objects for enough time due to frequent cache replacement which in turn results in a higher cache miss.

This very same explanation, however, contradicts with our observation in cases of large cache sizes which should have not resulted in poor prediction performance, since with larger cache size we are able to cache more objects for longer time period. Obviously, larger cache sizes are expected to result in a poor overall performance by requiring bigger memory space but this should have not been true if we are to consider prediction performance only. We plan to investigate this thoroughly in our future work alongside experimenting with multiple caching management algorithms.

In Figures 5-16, 2-18 and 5-26 it can be observed that the minimum performance of the scheme is higher than the minimum performance of all algorithms which implies that this algorithm is not directly correlated with any of the individual algorithms rather it is a combination of all. In other words, since the minimum prediction performance of the proposed framework is higher than all, we can conclude that the scheme is not shouldered by any single algorithm. This fact is reflected not only with minimum performance but maximum and average performances as well.

Our findings show that the proposed framework has a higher prediction performance which addresses the first of our two objectives stated earlier. The fact that the proposed framework is less aggressive than its counterparts directly addresses our second specific objective. Aggressive schemes consume resources extensively which hand held devices simply do not have. Hence, less aggressive schemes are obviously preferable in such contexts.

Since the process of prefetching and other related compute intensive tasks could be moved to a machine with larger computing power and memory in the request response chain without affecting the framework (e.g., a proxy server), this scheme eliminates the need for the client end device to shoulder these tasks solely.

In terms of generalization, even though we have not experimented with all contemporary prefetching schemes, we can safely conclude that the performance improvement will remain intact since our experimentation used similar data source and related algorithms to make sure the performance change observed is mainly due to the aggregation of algorithms. Of course, experimenting with more prefetching schemes is required which we plan to do in the future.

In terms of compatibility, the scheme can accommodate all existing prefetching schemes as it does not focus on any specific characteristics of any of these algorithms rather it makes use of a cross cutting property that exists in all contemporary schemes. Subsequently, any prefetching algorithm can be incorporated to this framework as long as it produces a list of candidate objects to prefetch. Figuring out which specific algorithms lead to maximum performance is yet to be worked on in our future work.

# Chapter 7 - Conclusion

The World Wide Web is progressing towards becoming the one stop no one escapes from to achieve anything in our daily customs. But latency has been and still is an issue that needs proper addressing to make users' experience even more pleasant. Caching was the first method towards addressing this hitch but the performance gain observed, while encouraging, is extremely limited. Prefetching was amended to caching to raise the performance gain due to caching techniques.

Contemporary works on prefetching are focused in two directions, namely, history based approaches and semantic/content based approaches. Both methods try to find useful patterns that could help in predicting user's future access. The difference in these two methods is the source they use to unveil the patterns. In semantic schemes, the content creator explicitly exposes the patterns while for the case of history based approaches, intelligent algorithms learn these patters through experience (i.e., from users past access history).

By their nature, history based schemes are not able to predict future access to web objects that were not accessed so far. This is because new objects do not appear in the access log used by the algorithm to learn from. In contrary, content based scheme could predict future access even if the objects were never accessed in the past since the pattern is revealed by the content creator. This makes content based schemes suitable for the dynamic web environment we are experiencing nowadays whereby web objects are created so fast that history based schemes simply do not get the time to see useful patterns.

Yet, currently, history based schemes generally outperform their content based counterparts mainly due to the lack of significantly enough semantic information embedded within today's web content. However, as more and more semantic information becomes available, there will be a need to move to semantic prefetching schemes to cope with the ever dynamic web environment we have.

Some researchers tried to combine these two methods of prefetching to benefit from the pros of each side. However, most of their works were directed in combining handpicked few algorithms from each direction of prefetching rather than forming a generic scheme that could be used to combine multiple algorithms as they become available. Having such framework enables incorporation of new algorithms that capture new context as they become available.

In this thesis work we have presented such a framework for multi-source prefetching with adaptive weight values to enable collaborative prefetching. We have made our evaluation focussing on prediction efficiency of the overall scheme and we have found encouraging results.

## 7.1 Contribution

The primary contribution of this work is in developing a novel framework that could accommodate a multitude of contemporary prefetching algorithms. Unlike other works, mentioned in the related work Chapter we did not set out to combine selected prefetching algorithms but we planned on creating a framework that can accommodate any one of the prediction algorithm that anticipate user's future access in the prefetching context.

Given the above stated need to move towards semantic prefetching schemes, this framework can serve as one possible solution in the transition from contemporarily pure history based prefetching technology towards fully or partially semantics driven prefetching schemes. This is possible because the framework allows any algorithm to participate in the prediction process without negatively affecting overall performance via the reputation based control mechanism that assigns weights to participating algorithms according to the performance exhibited by each. When we have semantic schemes that perform well enough to suppress their history based counterparts, the scheme will in time virtually turn off the history based approaches.

By including multiple prediction algorithms within the prefetching scheme we could capture application level relationships that help in accurately predicting user's future access in a dynamic environment through custom algorithms that make use of application level information.

Even though our work is mainly pertinent to prefetching, since the framework's critical component is directed to user's future access prediction via collaboration, it can be adopted to domains with similar requirements on anticipating user's future access.

## 7.2 Future Work

Due to time, computational resource and scope limitations we have not addressed a number of issues we deem important for the realization of the framework we have proposed. In the future we plan to address these issues. The first of such issues is a detailed study of the effect of the cache management policy adopted on the framework. We would also like to understand the unexpected result presented in the previous Chapter that shows an increase cache size to result in decreased overall performance.

Furthermore, we plan to conduct a detailed study on the reputation which served as an abstract weight value that captures the performance of each participating algorithm. We would like to expand this measure to include security related issues such as trustworthiness of each algorithm for the framework to be feasible in production environment. Implementing/Adopting and testing a complex context management mechanism is also an envisioned task as we have not done so in this work. Context management is a cross cutting challenge in various sub-disciplines of Computer Science but it is mainly addressed in pervasive computing which we hope to adopt a suitable solution from.

What is more, testing the framework with more algorithms from the various genres available in the prefetching world is also a planned task. Our implementation and evaluation focused on minimal configuration which led us to using only a handful of similar algorithms to magnify the performance gain observed due to the combination of algorithms. However, in future works we plan to experiment with a plethora of algorithms from both semantic and history based approaches.

We plan to conduct a trace driven evaluation to gauge the overall performance as opposed to the measurement we have conducted that evaluates only the performance of the prediction task. Integrating the framework to current web technology tools such as web browsers, proxy server software and web and application servers is also a task ahead.

# Declaration

I, Undersigned, declare that this thesis is my original work and has not been presented for degree in any other university, and that all sources of material used for the thesis have been acknowledged.

Declared by:

Name:        **<u>Yoseph Berhanu Alebachew</u>**

Signature:   ______________________

Date         ______________________

Confirmed by advisor:

Name:        **<u>Mulugeta Libsie (PhD)</u>**

Signature:   ______________________

Date         ______________________

Place and date of submission: Addis Ababa University, May 26, 2014.